\documentclass[11pt]{article}
\usepackage[dvips]{graphicx}
\usepackage{amssymb}
\usepackage{amsmath}
\usepackage{amssymb}
\usepackage{overcite}
\usepackage{epsf}
\input{epsf}
\begin{document}
\begin{titlepage}
\title{\bf  Electronic  Dynamics   of the Anderson Model. The Many-Body Approach}  
\author{A. L. Kuzemsky 
\\
{\it Bogoliubov Laboratory of Theoretical Physics,} \\
{\it  Joint Institute for Nuclear Research,}\\
{\it 141980 Dubna, Moscow Region, Russia.}\\
{\it E-mail:  kuzemsky@theor.jinr.ru} \\
{\it http://theor.jinr.ru/\symbol{126}kuzemsky}}
\date{}
\maketitle
\begin{abstract}
A review of  electronic dynamics of single-impurity and
many-impurity Anderson models is contained in this report.
Those models are used widely   for many of the applications in diverse fields of interest, such as surface physics,
theory of chemisorption and adsorbate reactions on metal surfaces, physics of intermediate valence
systems, theory of heavy fermions, physics of quantum dots and other nanostructures. While standard treatments are generally 
based on perturbation methods, our approach is based  on the non-perturbative technique for the   thermodynamic 
Green functions.  The method of the irreducible Green  functions  is used as the basic tool.
This irreducible Green  functions method allows one to describe the quasiparticle spectra
with damping of the strongly correlated electron systems  in a very general and natural way and to construct the relevant
dynamical solution in a self-consistent way on the level of Dyson equation without decoupling the chain of the equations
of motion for the Green functions.  The subject matter includes the improved interpolating solution of the Anderson model.
It was shown that an interpolating approximation, which simultaneously reproduces the weak-coupling
limit up to second order in the interaction strength U and the strong-coupling limit up to
second order in the hybridization V (and thus also fulfils the atomic limit) can be formulated
self-consistently. This approach offers a new way for the systematic construction of   approximate interpolation
dynamical  solutions of   strongly correlated electron systems.
\vspace{1cm}


\vspace{1cm}
%\noindent\textbf{PACS}:  
%
%%%%%%%%%%%%%%%%%%%%%%
%
\end{abstract}
\end{titlepage}
\newpage
\tableofcontents
\newpage
%
%
%
%
%
%
%
%%%%%%%%%%%%%%%%%%%%%%% 
\section{Introduction}
%%%%%%%%%%%%%%%%%%%%%%% 
%
%
%
In this review  we discuss the many-body quasiparticle dynamics of the
Anderson impurity model~\cite{ander61,and79}  and its generalizations~\cite{ander64}  in the framework
of the equation-of-motion method~\cite{tyab,kuz09,kuzapp76,kuzam91,kpb93,kuz94,ckw96,kuzem96,kuapp97} at finite 
temperatures.\\ 
The studies of strongly correlated electrons in solids and their quasiparticle dynamics  are intensively
explored subjects in solid state physics~\cite{kuz09,kuzem05,bech15,rubi00}. 
Electronic dynamics in the bulk and at the surface of solid materials are well known to play a key role in a variety of
physical and chemical phenomena~\cite{kuz09,rubi00,hoff88,kiko94,dona97,adam97,atki09}. One of  the main aspects of such studies is the interaction of 
low-energy electrons with solids, where the calculations of inelastic lifetimes of both low-energy electrons in 
bulk materials and image-potential states at metal surfaces are highly actual problems. The calculations of inelastic 
lifetimes was made as a rule in a model of the homogeneous electron gas~\cite{rubi00},
by using various approximate representations of the electronic response of the medium. Band-structure calculations, which
have been  carried out in literature may give a partial information only.\\ 
The band-structure approach~\cite{bech15,kahir03,mar04,huang08} suffers from well known limitations. It cannot be validated in full measure in the
case of very narrow bands and strongly correlated localized electrons~\cite{kuz09,kudr84,canad12}.
An alternative approach is connected
with using correlated fermion lattice models, like Anderson~\cite{ander61,and79} 
and Hubbard model~\cite{hubb1,hubb2,hubb3,hubb4,hubb5,hubb6,kuz78,mizia07,gors11}.\\
The principal importance of this approach is related with the \emph{dual  character} of electrons in  
dilute magnetic alloys~\cite{matt65,fis71,faul72,cole77,fis78,gavo81},
in transition metal oxides~\cite{kuz09,kuz00,qprot02,jgood97,zhou98,jgood00,good01,good14,capo09,veron10,wino14}, 
intermediate-valence solids~\cite{gerd86,rama90,zhu07}, heavy fermions~\cite{kuz94,gerd86,steg88,hew93,gehr02},
high-$T_{c}$ superconductors~\cite{kuz94,uchi15}, etc. In these materials electrons exhibit both
localized and delocalized features~\cite{kuz09,kuz00}. For example in paper~\cite{zhu07}
the electronic structure in solid phases of plutonium was discussed. The electrons in the outermost orbitals of 
plutonium show qualities of both \emph{atomic} and \emph{metallic} electrons. The metallic aspects of electrons and 
the electron duality that effect the electronic, magnetic and other properties of elements were manifested clearly.\\
The basic models to describe correlated electron systems are the single-impurity Anderson model
(SIAM)~\cite{ander61,and79,matt65,gehr02,rama94,ogur02}, periodic Anderson model (PAM)~\cite{mosca96} and the Hubbard model which exhibit the
key physical feature, i.e., the competition between kinetic energy
(itinerant) and potential energy (localized) effects~\cite{kuz09,kuz00,qprot02}. \\
Indeed, the standard approach which is valid mainly for the simple and noble metals is provided by the band theory formalism 
for the calculation of the electronic structure of solids. For a better understanding of how structure and properties of 
solids may be related the chemically insightful concept of orbital interaction and the essential machinery of band theory should be taken
into account~\cite{canad12} to  reveal links between the crystal and electronic structure of periodic systems. In such a way, 
it was  possible shown~\cite{canad12} how important tools for understanding properties of solids like the density of states, 
the Fermi surface etc.,  can be qualitatively formulated and used   to rationalize experimental observations. 
It was shown that extensive use of the orbital interaction approach appears to be a very efficient way of building bridges 
between physically and chemically based notions to understand the structure and properties of solids.\\
The Anderson and Hubbard models found a lot of applications in studies of surface physics~\cite{hoff88,adam97},  theory of chemisorption 
and adsorption~\cite{yang96,bell76,bren78,gfthe06,davyd12} and various aspects of physics of 
quantum dots~\cite{brod99,kouw01,mann02,pete04,ashok01,nazar02,kuo03,ashok07,stef09,citro09,ashok12,kuca14}.
However in spite of many theoretical efforts  a fully satisfactory
solution of the dynamical problem is still missing. The Bethe-ansatz
solution of the SIAM  allows for the determination of the ground state and thermodynamic static properties, 
but it does not allow for a determination of the dynamical properties. For their understanding the
development of improved and reliable approximations is still justified and desirable. 
In this context it is of interest to consider an  interpolating and improved interpolating approximations
which were proposed in the  papers~\cite{kuzapp76,kuzam91,kpb93,kuz94,ckw96,kuzem96,kuapp97}.
We will show that a self-consistent  approximation for the SIAM
can be formulated which reproduces all relevant exactly solvable limits and
interpolates between the strong- and  weak-coupling limit.\\
In connection with the dynamical properties the one-particle Green
function is the basic quantity to be calculated. 
Subject of  this survey is primarily devoted to the analysis of the
relevant many-body dynamic  solution  of the single impurity Anderson model  
and its \emph{correct functional structure}. We wish to emphasize that the
correct functional structure actually arises  both from the self-consistent many-body approach
and intrinsic nature of the model itself. The important representative quantity
is the \emph{spectral intensity} of the Green function  at low energy and low temperature. 
Hence, it is desirable to have a consistent and closed analytic representation
for the one-particle Green function of  SIAM. \ 
The papers~\cite{kuzapp76,kuzam91,kpb93,kuz94,ckw96,kuzem96,kuapp97} clearly show  the importance of the calculation
of the Green function  and spectral densities for SIAM and the many-impurity Anderson model in a self-consistent
way. \\ 
In this terse overview the problem of consistent analytic  description of the
many-body dynamics of SIAM is analyzed in the framework of the
equation-of-motion approach for double-time thermodynamic Green functions~\cite{tyab,kuz09}. In
addition to the  
irreducible Green functions (IGF) approach~\cite{kuz78,igf89,kuznc94,kuzmpl97,kuzrnc02,kuzcmp10,kuzsf99,kuz04,dms05}, 
we use a new \textbf{exact identity}~\cite{kuzam91,kuzem96,kuzrnc02}, relating the one-particle and many-particle Green functions. 
Using this identity, it was possible to formulate a consistent and general scheme for
construction of generalized solutions of the Anderson model.  A new approach
for the complex expansion for the single-particle propagator in
terms of Coulomb repulsion $U$ and hybridization $V$ is discussed as well.
Using the exact identity, an essentially new many-body dynamic solution of SIAM was derived. 
%
%
%
%%%%%%%%%%%%%%%%%%%%%%%%%%%%%%%%%%% 
\section{Hamiltonian of the Models}
%%%%%%%%%%%%%%%%%%%%%%%%%%%%%%%%%%% 
%
%
\subsection{Single-impurity Anderson model (SIAM)}
%(i) \textbf{Single-impurity Anderson model (SIAM)}.\\
%
%\vspace{1cm}
%
The Hamiltonian of the SIAM can be written in the form
\begin{eqnarray}
\label{siam}
H = \sum_{\vec k\sigma}\epsilon_{\vec k}c^{\dagger}_{\vec k\sigma}c_{\vec k\sigma} + \sum_{\sigma}
E_{0\sigma}f^{\dagger}_{0\sigma}f_{0\sigma} + \\ \nonumber
\frac{U}{2}\sum_{\sigma}f^{\dagger}_{0\sigma}f_{0\sigma}f^{\dagger}_{0-\sigma}f_{0-\sigma} +
 \sum_{\vec k\sigma} V_{\vec k}  (c^{\dagger}_{\vec k\sigma}f_{0\sigma} +
f^{\dagger}_{0\sigma}c_{\vec k \sigma}),
\end{eqnarray}
where $c^{\dagger}_{\vec k\sigma}$ and $f^{\dagger}_{0\sigma}$ are the
creation operators for conduction and localized electrons;
$\epsilon_{\vec k}$ is the conduction electron dispersion,
$E_{0\sigma}$ is the
localized (f-) electron energy level and $U$  is the intra-atomic Coulomb
interaction at the impurity site. $V_{\vec k}$ represents the $s-f$
hybridization. In the following consideration we will omit the vector notation for the sake of brevity. 
%
%(ii) \textbf{Periodic Anderson Model (PAM)}.\\
\subsection{Periodic Anderson Model (PAM)}
Let us now consider  a lattice generalization of SIAM,  the
so-called periodic Anderson model (PAM). The basic assumption of
the periodic impurity Anderson model is the presence of two
well-defined subsystems,   i.e.  the Fermi sea of nearly free
conduction electrons and the localized impurity orbitals embedded
into the  continuum of conduction electron states (in rare-earth
compounds, for instance, the continuum is actually a mixture of
$s$, $p$, and $d$ states, and the localized orbitals are $f$
states). The simplest form of PAM
\begin{eqnarray}
\label{eq.59}
 H =\sum_{k\sigma}\epsilon_{k}c^{\dagger}_{k\sigma}c_{k\sigma} +
\sum_{i\sigma} E_{0} f^{\dagger}_{i\sigma}f_{i\sigma} +
U/2\sum_{i\sigma}n_{i\sigma}n_{i-\sigma} + \\
\nonumber \frac {V}{\sqrt N} \sum_{ik\sigma} ( exp(i
kR_{i})c^{\dagger}_{k\sigma}f_{i\sigma} + exp (-ikR_{i})
f^{\dagger}_{i\sigma}c_{k\sigma})
\end{eqnarray}
assumes a one-electron energy level $E_{0}$,  hybridization
interaction $V$, and the Coulomb interaction $U$ at each lattice
site. Using the transformation
\begin{equation}
 c^{\dagger}_{k\sigma} = \frac {1}{\sqrt N} \sum_{j}  exp(-i
kR_{j}) c^{\dagger}_{j\sigma}; \quad c_{k\sigma} = \frac
{1}{\sqrt N} \sum_{j}  exp(ikR_{j})c_{j\sigma} 
\end{equation}
 the Hamiltonian
(\ref{eq.59}) can be rewritten in the Wannier representation:
\begin{eqnarray}
\label{eq.60}
 H =\sum_{ij\sigma} t_{ij}c^{\dagger}_{i\sigma}c_{j\sigma} +
\sum_{i\sigma} E_{0} f^{\dagger}_{i\sigma}f_{i\sigma} +
U/2\sum_{i\sigma}n_{i\sigma}n_{i-\sigma} + \\
\nonumber V \sum_{i\sigma} ( c^{\dagger}_{i\sigma}f_{i\sigma} +
f^{\dagger}_{i\sigma}c_{i\sigma}).
\end{eqnarray}
If one retains the $k$-dependence of the hybridization matrix
element $V_{k}$ in (\ref{eq.60}), the last term in the PAM
Hamiltonian  describing the hybridization interaction between the
localized impurity states and extended conduction states and
containing the essence of a specificity  of the Anderson model,
is as follows
\begin{equation}
\sum_{ij\sigma} V_{ij}( c^{\dagger}_{i\sigma}f_{i\sigma} +
f^{\dagger}_{i\sigma}c_{i\sigma}); \quad V_{ij} = \frac {1}{ N}
\sum_{k} V_{k} exp[ik(R_{j} - R_{i})].  
\end{equation}
The on-site hybridization $V_{ii}$ is equal to zero for symmetry reasons.  
Hence  the Hamiltonian of PAM in the Bloch representation  takes the form
\begin{eqnarray}
\label{eq.61}
 H =\sum_{k\sigma}\epsilon_{k}c^{\dagger}_{k\sigma}c_{k\sigma} +
\sum_{i\sigma} E_{k} f^{\dagger}_{k\sigma}f_{k\sigma} +
U/2\sum_{i\sigma}n_{i\sigma}n_{i-\sigma} + \\
\nonumber  \sum_{k\sigma} V_{k} (
c^{\dagger}_{k\sigma}f_{k\sigma} +
 f^{\dagger}_{k\sigma}c_{k\sigma}).
\end{eqnarray}
Note that as compared to the SIAM, the PAM has its own specific
features. This can lead to peculiar magnetic properties for
concentrated rare-earth systems where the criterion for magnetic
ordering depends on the competition between indirect RKKY-type
interaction\cite{kuzem05} (not included into SIAM) and the
Kondo-type singlet-site screening (contained in SIAM).\\
The inclusion of inter-impurity correlations makes the problem even more
difficult. Since these inter-impurity effects play an essential
role in physical behaviour of real
systems~\cite{kpb93,kuzrnc02}, it is instructive to consider
the two-impurity Anderson model (TIAM) too. 
\subsection{Two-Impurity Anderson Model (TIAM)}
%
%(iii) \textbf{Two-impurity Anderson model (TIAM)}.\\
%
The two-impurity Anderson model was  considered by Alexander and
Anderson~\cite{ander64}.  They  put forward a theory which introduces
the impurity-impurity interaction within a game of parameters.\\
The Hamiltonian of TIAM reads
\begin{eqnarray}
\label{eq.62}
 H =\sum_{ij\sigma} t_{ij}c^{\dagger}_{i\sigma}c_{j\sigma} +
\sum_{i=1,2\sigma} E_{0i} f^{\dagger}_{i\sigma}f_{i\sigma} +
U/2\sum_{i=1,2\sigma}n_{i\sigma}n_{i-\sigma} + \\
\nonumber  \sum_{i\sigma} (
V_{ki}c^{\dagger}_{i\sigma}f_{i\sigma} +
V_{ik}f^{\dagger}_{i\sigma}c_{i\sigma}) + \sum_{\sigma} (
V_{12}f^{\dagger}_{1\sigma}f_{2\sigma} +
V_{21}f^{\dagger}_{2\sigma}f_{1\sigma})
\end{eqnarray}
where  $E_{0i}$ are the position energies of  localized states (for simplicity, we consider identical impurities and $s$-type
i.e.  non-degenerate) orbitals: $E_{01} = E_{02} = E_{0}$.
The  hybridization matrix element $V_{ik}$ was
discussed in in detail in Ref.~\cite{kpb93}  As for the TIAM, the situation with the
right definition of the parameters  $V_{12}$ and $V_{ik}$ is not
very clear. The definition of $V_{12}$ in~\cite{ander64} is the
following:
\begin{equation}
V_{12} = V^{\dagger}_{21} = \int
\phi^{\dagger}_{1}(\vec r) H_{f}\phi_{2}(\vec r)  dr. 
\end{equation}
Note that  $H_{f}$ is without "H-F" (Hartree-Fock) mark. The essentially local character of
the Hamiltonian $H_{f}$ clearly shows that $V_{12}$ describes the
direct coupling between nearest neighboring sites (for a
detailed discussion see Ref.\cite{kpb93} where the hierarchy of the Anderson models was  discussed too).
%
% 
% 
%%%%%%%%%%%%%%%%%%%%%%%%%%%%%%%%%%%%%%%%%%%%%%%%%%%%%%%%%
 \section{The Method of Irreducible Green  Functions}     
%%%%%%%%%%%%%%%%%%%%%%%%%%%%%%%%%%%%%%%%%%%%%%%%%%%%%%%%% 
%
When working with infinite hierarchies of equations
for Green functions~\cite{tyab,kuz09} the main problem is finding the methods for
their efficient decoupling, with the aim of obtaining a
closed system of equations, which determine the Green functions. In the 
papers~\cite{kuz78,igf89,kuznc94,kuzmpl97,kuzrnc02}
devoted to studies of lattice systems of interacting fermions it was shown that for a
wide range of problems in statistical mechanics~\cite{kuzcmp10,kuzsf99,kuz04,dms05} and
theory of condensed matter one can outline a fairly systematic
recipe for constructing approximate solutions
in the framework of \emph{irreducible Green  functions}
method. Within this approach one can look from a unified
point of view at the main problems of fundamental
characters arising in the method of two-time temperature
Green functions. \\ 
The method of irreducible Green  functions is
a useful reformulation of the ordinary Bogoliubov-Tyablikov method of equations of motion~\cite{tyab,kuz09}. 
The constructive idea can be summarized as follows. During
calculations of single-particle characteristics of the system
(the spectrum of quasiparticle excitations, the density
of states, and others) it is convenient to begin from
writing down Green function  as a formal solution of the Dyson
equation. This will allow one to perform the necessary
decoupling of many-particle correlation functions in
the mass operator. This way one can to control the
decoupling procedure conditionally, by analogy with
the diagrammatic approach.  In this
approach the infinite hierarchy of coupled equations for
correlation functions is reduced to a few relatively simple
equations that effectively take into account the
essential information on the system under consideration,
which determine the special features of this concrete
problem. \\ 
It is necessary to stress that the structure
of solutions obtained in the framework of irreducible
Green functions method is very sensitive to the order of equations
for Green functions~\cite{kuzrnc02} in which irreducible parts are separated.
This in turn determines the character of the approximate
solutions constructed on the basis of the exact representation.\\
Now we introduce the retarded, advanced, and causal Green function:
\begin{eqnarray}
\label{e39} G^{r}(A,B; t-t') = \langle \langle A(t), B(t')\rangle \rangle^{r} = -i\theta(t -
t')\langle [A(t),B(t')]_{\eta}\rangle, \,  \eta = \pm,
 \\ \label{e40}
G^{a}(A,B; t-t') = \langle \langle A(t), B(t')\rangle \rangle^{a} =
i\theta(t' - t)\langle [A(t),B(t')]_{\eta}\rangle, \, \eta = \pm,  \\
G^{c}(A,B; t-t') = \langle \langle A(t), B(t')\rangle \rangle^{c} = i T \langle A(t)B(t')\rangle =   \\ \nonumber
i\theta(t - t')\langle A(t)B(t')\rangle + \eta i\theta (t'- t) \langle B(t')A(t)\rangle, \,
\eta = \pm. \label{e41}
\end{eqnarray}
Here $\langle \ldots \rangle$  is the average over the grand canonical ensemble,
$\theta(t)$ is the Heaviside step function;  the
square brackets denote either commutator or anticommutator
$(\eta = \pm)$:
\begin{equation} \label{e42}
[A,B]_{- \eta} = AB - \eta BA.
\end{equation}
An important ingredient for Green function application is their
temporal evolution. In order to derive the corresponding
evolution's equation, one has to differentiate Green function
over one of its arguments.
In order to clarify the above general description, let
us consider the equations of motion   for the
retarded Green function    of the form $ \langle \langle A(t), A^{\dagger}(t')\rangle \rangle $ 
\begin{equation}
\label{e70} \omega G(\omega) = \langle [A, A^{\dagger}]_{\eta} \rangle + \langle \langle [A,
H]_{-} \vert A^{\dagger}\rangle \rangle_{\omega}. 
\end{equation}
The irreducible (\textbf{ir}) Green function is defined by
\begin{equation}
\label{e71} ^{(ir)}\langle \langle[A, H]_{-}\vert A^{\dagger}\rangle \rangle = \langle \langle [A, H]_{-}
- zA\vert A^{\dagger}\rangle \rangle.
\end{equation}
The unknown constant $z$ is found from the condition
\begin{equation}
\label{e72} \langle [^{(ir)}[A, H]_{-}, A^{\dagger}]_{\eta} \rangle = 0.
\end{equation}
It is worth noting  that instead of finding
the irreducible part of Green function 
\begin{equation}
(^{(ir)}\langle \langle [A, H]_{-}\vert A^{\dagger}\rangle \rangle),
\end{equation}
one can absolutely equivalently consider \emph{the irreducible operators}
\begin{equation}
(^{(ir)}[A, H]_{-}) \equiv ([A, H]_{-})^{(ir)}. 
\end{equation}
Therefore, we will use both the notation ($ ^{(ir)}\langle \langle A  \vert B \rangle \rangle$) and $\langle \langle (A)^{(ir)}  \vert B \rangle \rangle$), whichever
is more convenient and compact. Equation (\ref{e72})
implies
\begin{equation}
\label{e73} z = \frac{\langle [[A, H]_{-}, A^{\dagger}]_{\eta} \rangle}{\langle [A,
A^{\dagger}]_{\eta}\rangle} =
 \frac{M_{1}}{M_{0}}.
\end{equation}
Here, $M_{0}$ and $M_{1}$   are the zero and first moments of the
spectral density~\cite{tyab,kuz09}. Green  function is
called \textbf{irreducible} (i.e. impossible to reduce to a desired, simpler, or smaller form or amount) if it cannot 
be turned into a lower order Green function via decoupling. The well-known objects in
statistical physics are irreducible correlation functions. In the framework of the diagram
technique the irreducible vertices are a set of
graphs, which cannot be cut along a single line. The
definition (\ref{e71}) translates these notions to the language
of retarded and advanced Green  functions. We
attribute all the mean-field renormalizations that are
separated by Eq. (\ref{e71}) to Green function within a \emph{generalized mean field}
approximation
\begin{equation}
\label{e74} G^{0}(\omega) = \frac{\langle [A,
A^{\dagger}]_{\eta}\rangle}{(\omega - z)}.
\end{equation}
For calculating Green function (\ref{e71}), $\,  ^{(ir)}\langle \langle [A, H]_{-}(t),A^{\dagger}(t')\rangle \rangle,$   we
make use of differentiation over the second time $t'$.
Analogously to Eq. (\ref{e71}) we separate the irreducible
part from the obtained equation and find
\begin{equation} \label{e75} 
G(\omega) = G^{0}(\omega) + G^{0}(\omega) P(\omega) G^{0}(\omega).
\end{equation}
Here, we introduced the scattering operator
\begin{equation}
\label{e76} P = (M_{0})^{-1}\biggl (\langle \langle([A,
H]_{-})^{(ir)}   \vert ([A^{\dagger}, H]_{-})^{(ir)} \rangle \rangle \biggr ) (M_{0})^{-1}.
\end{equation}
In complete analogy with the diagram technique one
can use the structure of Eq. (\ref{e75}) to define the mass
operator $M$:
\begin{equation} \label{e77} P = M + M G^{0} P.
\end{equation}
As a result we obtain the exact Dyson equation (we
did not perform any decoupling yet) for two-time temperature
Green functions:
\begin{equation} \label{e78} G =
G^{0} + G^{0} M G.
\end{equation}
According to Eq.(\ref{e77}), the mass operator $M$ (also
known as the self-energy operator) can be expressed in
terms of the \textbf{proper} (called \emph{connected} within the diagram
technique) part of the many-particle irreducible
Green function. This operator describes inelastic scattering processes,
which lead to damping and to additional renormalization
of the frequency of self-consistent quasiparticle
excitations. One has to note that there is quite a
subtle distinction between the operators $P$ and $M$. Both
operators are solutions of two different integral equations
given by Eqs. (\ref{e77}) and (\ref{e78}), respectively. However,
only the Dyson equation (\ref{e78}) allows one to write
down the following formal solution for the Green function:
\begin{equation}
\label{e79} G = [ (G^{0})^{-1} - M ]^{-1}.
\end{equation}
This fundamental relationship can be considered as
an alternative form of the Dyson equation, and as the
\textbf{definition} of the mass operator under the condition that
the Green function within the generalized mean-field approximation,
$G^{0}$, was appropriately defined using the equation
\begin{equation}
\label{e80} G^{0}G^{-1} + G^{0}M = 1.
\end{equation}
In contrast, the operator $P$ does not satisfy Eq. (\ref{e80}).
Instead we have
\begin{equation}
\label{e81} (G^{0})^{-1} - G^{-1} = P G^{0}G^{-1}.
\end{equation}
Thus, it is \textbf{the functional structure} of Eq. (\ref{e79}) that
determines the essential differences between the operators
$P$ and $M$. To be absolutely precise, the definition
(\ref{e77}) has a symbolic character. It is assumed there that
due to the similar structure of equations (\ref{e39}) - (\ref{e41})
defining all three types of Green functions, one can use the causal Green functions
at all stages of calculation, thus confirming the sensibility
of the definition (\ref{e77}). Therefore, one should rather
use the phrase "an analogue of the Dyson equation".
Below we will omit this stipulation, because it will not
lead to misunderstandings. One has to stress that the
above definition of irreducible parts of the Green function (irreducible
operators) is nothing but a general scheme. The
specific way of introducing the irreducible parts of the
Green function depends on the concrete form of the operator $A$ on
the type of the Hamiltonian, and on the problem under
investigation. \\ 
Thus, we managed to reduce the derivation
of the complete Green function to calculation of the Green function in the
generalized mean-field approximation and with the
generalized mass operator. The essential part of the
above approach is that the approximate solutions are
constructed not via decoupling of the equation-of-motion
hierarchy, but via choosing the functional form
of the mass operator in an appropriate self-consistent
form. That is, by looking for approximations of the
form $ M \approx F[G]$. Note that the exact functional structure
of the one-particle Green function (\ref{e79}) is preserved in this
approach, which is quite an essential advantage in comparison
to the standard decoupling schemes.
%
%
%%%%%%%%%%%%%%%%%%%%%%%%%%%%%%%%%%%%%%%%%%%%%%%%%%%%%%%%%%% 
\section{The Irreducible Green Functions Method and SIAM}
%%%%%%%%%%%%%%%%%%%%%%%%%%%%%%%%%%%%%%%%%%%%%%%%%%%%%%%%%% 
%
%
After discussing some of the basic facts about the correct
functional structure of the relevant dynamic solution of
correlated electron models we are looking for, described in
previous Chapter, we   give a similar consideration for SIAM. It
was shown in Refs.~\cite{kuzam91,ckw96,kuzem96,kpb93}, using the minimal algebra of relevant
operators, that the construction of the generalized mean fields for SIAM is quite
nontrivial for the strongly correlated case, and it is rather
difficult to get it from an intuitive physical point of view.\\ 
To proceed let us consider first the following matrix Green function
\begin{equation}
\label{eq.224}
 \hat G (\omega) =
\begin{pmatrix}
\langle \langle c_{k\sigma}\vert c^{\dagger}_{k\sigma} \rangle \rangle &\langle \langle c_{k\sigma}\vert f^{\dagger}_{0\sigma} \rangle \rangle \\ 
\langle \langle f_{0\sigma}\vert c^{\dagger}_{k\sigma} \rangle \rangle & \langle \langle f_{0\sigma}\vert f^{\dagger}_{0\sigma} \rangle \rangle \\
\end{pmatrix}. 
\end{equation}
Performing the first-time differentiation and defining the irreducible Green function
\begin{eqnarray}
\label{eq.225}
(^{(ir)}\langle \langle f_{0\sigma}f^{\dagger}_{0-\sigma}f_{0-\sigma} \vert
f^{\dagger}_{0\sigma} \rangle \rangle_ {\omega})  =
 \langle \langle f_{0\sigma}f^{\dagger}_{0-\sigma}f_{0-\sigma}\vert
f^{\dagger}_{0\sigma} \rangle \rangle_{\omega} - \\ \nonumber - \langle n_{0-\sigma} \rangle
\langle \langle f_{0\sigma} \vert f^{\dagger}_{0\sigma} \rangle \rangle_{\omega},
\end{eqnarray}
we obtain the following equation of motion in the matrix form
\begin{equation}
\label{eq.226} \sum_{p} \hat F_{p}(\omega ) \hat G_{p} (\omega) =
\hat 1 + U \hat D^{(ir)} (\omega),
\end{equation}
where all definitions are rather evident. Proceeding  further
with the IGF technique, the equation of motion (\ref{eq.226}) may be
 rewritten exactly in the form of the Dyson equation
\begin{equation}
\label{eq.227} \hat G (\omega) = \hat G^{0}(\omega) +  \hat
G^{0}(\omega) \hat M( \omega) \hat G (\omega).
\end{equation}
The generalized mean field Green function $G^{0}$  is defined by
\begin{equation}
\label{eq.228} \sum_{p} F_{p}(\omega) G_{p}^{0} (\omega) = \hat I.
\end{equation}
The explicit solutions for diagonal elements of $G^{0}$ are
\begin{eqnarray}
\label{eq.229} \langle \langle f_{0\sigma} \vert f^{\dagger}_{0\sigma} \rangle \rangle^{0}_
{\omega} = \Bigl ( \omega -E_{0\sigma} - Un_{-\sigma} -
S (\omega)) \Bigr )^{-1} \\
\label{eq.230} \langle \langle c_{k\sigma} \vert
c^{\dagger}_{k\sigma} \rangle \rangle^{0}_{\omega} = \Bigl ( \omega
-\epsilon_{k} - \frac { |V_{k}|^{2}}{\omega - E_{0\sigma} -
Un_{-\sigma} } \Bigr )^{-1},
\end{eqnarray}
where
\begin{equation}
\label{eq.231} S(\omega) = \sum_{k} \frac { |V_{k}|^{2}}{\omega - \epsilon_{k}}.
\end{equation}
The mass or self-energy  operator, which describes inelastic scattering processes, has the
following matrix form
\begin{equation}
\label{eq.232}
 \hat M (\omega) =
\begin{pmatrix} 
 0&0\\ 
 0&M_{0\sigma}\\
\end{pmatrix}, 
\end{equation}
where
\begin{equation}
\label{eq.233} M_{0\sigma} = U ^{2}  \Bigl ({} ^{(ir)}\langle \langle
f_{0\sigma}n_{0-\sigma} | f^{\dagger}_{0\sigma}n_{0-\sigma}
\rangle \rangle^{(ir)}_{\omega} \Bigr )^{(p)}.
\end{equation}
From the formal solution  of the Dyson equation (\ref{eq.227}) one obtains
\begin{eqnarray}
\label{eq.234} \langle \langle f_{0\sigma} \vert f^{\dagger}_{0\sigma} \rangle \rangle_
{\omega} =
 \Bigl ( \omega -E_{0\sigma} - Un_{-\sigma} - M_{0\sigma} -
S(\omega)  \Bigr )^{-1} \\
\label{eq.235} \langle \langle c_{k\sigma} \vert
c^{\dagger}_{k\sigma}\rangle \rangle_{\omega} = \Bigl ( \omega -\epsilon_{k} -
\frac { |V_{k}|^{2}}{\omega - E_{0\sigma} - Un_{-\sigma}-
M_{0\sigma}} \Bigr )^{-1}.
\end{eqnarray}
To calculate the self-energy in a self-consistent way, we have to
approximate it by lower-order Green functions. Let us start by
analogy with the Hubbard model with a pair-type approximation~\cite{kuz78,kuznc94,kuzrnc02}
\begin{align}
\label{eq.236} M_{0\sigma}(\omega) =  U^2 \int
\frac{d\omega_{1}d\omega_{2}d\omega_{3}}{\omega + \omega_{1} -
\omega_{2} - \omega_{3}} \times \\
~[n(\omega_{2})n(\omega_{3}) + n(\omega_{1})(1 - n(\omega_{2}) -
n(\omega_{3}))]
g_{0-\sigma}(\omega_{1})g_{0\sigma}(\omega_{2})g_{0-\sigma}(\omega_{3}),
\nonumber
\end{align}
where we   used the notation  
\begin{equation}
g_{0\sigma}(\omega) = -{1 \over \pi} \textrm{Im} \, \langle \langle f_{0\sigma} | f^{\dagger}_{0\sigma} \rangle \rangle_{\omega}. 
\end{equation}
The equations (\ref{eq.227}) and (\ref{eq.236}) constitute a closed
self-consistent system of equations for the single-electron Green function
for SIAM model, but only for weakly correlated  case. In
principle, we can use, on the r.h.s. of Eq.(\ref{eq.236}), any
workable first iteration-step form of the Green function and find a solution
by repeated  iteration. If we take for the first iteration step
the expression
\begin{equation}
\label{eq.237} g_{0\sigma}(\omega) \approx \delta(\omega -
E_{0\sigma} - Un_{-\sigma}),
\end{equation}
we get, for the self-energy, the explicit expression
\begin{eqnarray}
\label{eq.238} M_{0\sigma}(\omega) = U^{2} \frac {n(E_{0\sigma} +
Un_{-\sigma})(1 - n(E_{0\sigma} + Un_{-\sigma}))}{\omega -
E_{0\sigma} - Un_{-\sigma}} \\ = U^{2} Q_{-\sigma}(1 - Q_{-\sigma})G^{0}_{\sigma}(\omega),\nonumber
\end{eqnarray}
where 
\begin{equation}
  Q_{-\sigma} = n(E_{0\sigma} + Un_{-\sigma}), \quad  n(E) = \{\exp[(E - \mu)/k_BT]+1\}^{-1}. 
\end{equation}
This is the well-known  \emph{atomic  limit} of the self-energy~\cite{kuzam91,ckw96,kuzem96,kpb93}.\\ 
Let us try again another type of the approximation for $M$. The approximation which
we will use reflects the interference between the one-particle
branch and the collective one
\begin{eqnarray}
\label{eq.239}
\langle f_{0\sigma}(t)f^{\dagger}_{0-\sigma}(t)f_{0-\sigma}(t)
f^{\dagger}_{0-\sigma}f_{0-\sigma}f^{\dagger}_{0\sigma} \rangle^{(ir)} \approx \nonumber\\
\langle f^{\dagger}_{0\sigma}(t)f_{0\sigma} \rangle \langle n_{0-\sigma}(t)n_{0-\sigma} \rangle + \nonumber\\
\langle f^{\dagger}_{0-\sigma}(t)f_{0-\sigma} \rangle \langle f^{\dagger}_{0-\sigma}(t)f_{0\sigma}(t)
f^{\dagger}_{0\sigma}f_{0-\sigma} \rangle +\nonumber \\
\langle f^{\dagger}_{0-\sigma}(t)f_{0-\sigma} \rangle \langle f_{0-\sigma}(t)f_{0\sigma}(t)
f^{\dagger}_{0\sigma}f^{\dagger}_{0-\sigma} \rangle.
\end{eqnarray}
If we retain  only the first term  in (\ref{eq.239}) and make use
of the same iteration as in (\ref{eq.237}), we obtain
\begin{equation}
\label{eq.240} M_{0\sigma}(\omega) \approx U^{2} \frac {(1 -
n(E_{0\sigma} + Un_{-\sigma}))}{\omega - E_{0\sigma} -
Un_{-\sigma}} \langle n_{0-\sigma} n_{0-\sigma} \rangle.
\end{equation}
If we retain the second term in (\ref{eq.239}), we obtain
\begin{eqnarray}
\label{eq.241} M_{0\sigma}(\omega) = U^2 \int_{-\infty}^{+\infty}
d\omega_{1}d{\omega}_{2}\frac{1 + N(\omega_{1}) - n(\omega_{2})}
{\omega - \omega_{1} - \omega_{2}} \times \nonumber\\
\Bigl( -{1 \over \pi}\textrm{Im} \, \langle \langle S^{\pm}_{0} \vert S^{\mp}_{0} \rangle \rangle_{\omega_{1}} \Bigr ) 
\Bigl(-{1 \over \pi} \textrm{Im} \,  \langle \langle f_{0\sigma}\vert
f^{\dagger}_{0\sigma}\rangle \rangle_{\omega_{2}} \Bigr),
\end{eqnarray}
where the following notation were used:
\begin{equation}
 S^{+}_{0} = f^{\dagger}_{0\uparrow}f_{0\downarrow}; \quad S^{-}_{0} = f^{\dagger}_{0\downarrow}f_{0\uparrow}.
\end{equation}
It is
possible now  to rewrite (\ref{eq.241}) in a more convenient way 
\begin{align}
\label{eq.242} M_{0\sigma}(\omega) = U^2 \int d\omega'
\Bigl ( \cot \frac{\omega - \omega'}{2T} + \tan \frac{\omega'}{2T} \Bigr )
\Bigl(- \frac{1}{\pi} \textrm{Im} \, \chi^{\mp \pm} (\omega -
\omega')g_{0\sigma} (\omega') \Bigr).
\end{align}
The equations (\ref{eq.227}) and (\ref{eq.242}) constitute a
self-consistent system of equations for the single-particle Green function of
SIAM. Note  that spin-up and spin-down electrons are correlated
when they occupy the impurity level. So, this really improves the
standard mean-field theory in which just these correlations were missed. The role
of electron-electron correlation becomes much more crucial for the
case of strong correlation.
%
%%%%%%%%%%%%%%%%%%%%%%%%%%%%%%%%%%%%%%%%%%%
\section{SIAM. Strong Correlation}
%%%%%%%%%%%%%%%%%%%%%%%%%%%%%%%%%%%%%%%%%%%
%
The simplest relevant algebra of the operators used for the
description of the strong correlation has a similar form as for
that of the Hubbard model~\cite{hubb3,kuz78,kuznc94,kuzrnc02}. Let us represent the
matrix Green function (\ref{eq.224}) in the following form
\begin{equation}
\label{eq.243} \hat G (\omega) = \sum_{\alpha \beta} 
\begin{pmatrix}
\langle \langle c_{k\sigma}\vert c^{\dagger}_{k\sigma} \rangle \rangle & \langle \langle c_{k\sigma}\vert d^{\dagger}_{0 \beta \sigma}\rangle \rangle \\ 
\langle \langle d_{0 \alpha \sigma}\vert c^{\dagger}_{k\sigma}\rangle \rangle & \langle \langle d_{0\alpha \sigma}\vert d^{\dagger}_{0 \beta \sigma}\rangle \rangle \\
\end{pmatrix}.
\end{equation}
Here the operators $d_{0\alpha \sigma}$ and $d^{\dagger}_{0 \beta \sigma}$ are
\begin{eqnarray}
\label{eq.154} d_{i\alpha\sigma} =
n^{\alpha}_{i-\sigma}a_{i\sigma}, (\alpha = \pm);\quad
n^{+}_{i\sigma} = n_{i\sigma},\quad n^{-}_{i\sigma} = (1 -
n_{i\sigma});
\nonumber\\
\sum n^{\alpha}_{i\sigma} = 1; \quad
n^{\alpha}_{i\sigma}n^{\beta}_{i\sigma} =
\delta_{\alpha\beta}n^{\alpha}_{i\sigma}; \quad \sum_{\alpha}
d_{i\alpha\sigma} = a_{i\sigma}.
\end{eqnarray}
The new operators $d_{i\alpha \sigma}$ and $d^{\dagger}_{j\beta
\sigma}$ have complicated commutation rules, namely,
\begin{equation}
[d_{i\alpha \sigma}, d^{\dagger}_{j\beta \sigma}]_{+} = \delta_{ij}
\delta_{\alpha \beta}n^{\alpha}_{i-\sigma}. 
\end{equation}
Then we proceed by analogy with the calculations for the Hubbard
model. The equation of motion for the auxiliary matrix Green function
\begin{eqnarray}
\label{eq.244} \hat F_{\sigma}(\omega) =  
\begin{pmatrix}
\langle \langle c_{k\sigma}\vert c^{\dagger}_{k\sigma}\rangle \rangle & \langle \langle c_{k\sigma}\vert d^{\dagger}_{0 + \sigma}\rangle \rangle &\langle \langle c_{k\sigma} \vert d^{\dagger}_{0-\sigma}\rangle \rangle\\ 
\langle \langle d_{0 + \sigma}\vert c^{\dagger}_{k\sigma}\rangle \rangle & \langle \langle d_{0 + \sigma}\vert d^{\dagger}_{0 + \sigma}\rangle \rangle& \langle \langle d_{0+\sigma} \vert d^{\dagger}_{0-\sigma}\rangle \rangle \\
\langle \langle d_{0-\sigma} \vert c^{\dagger}_{k\sigma}\rangle \rangle& \langle \langle d_{0-\sigma} \vert d^{\dagger}_{0+\sigma}\rangle \rangle&\langle \langle d_{0-\sigma} \vert d^{\dagger}_{0-\sigma}\rangle \rangle\\ \nonumber
\end{pmatrix}
\end{eqnarray}
is of the following form
\begin{equation}
\label{eq.245} \hat E \hat F_{\sigma}(\omega) - \hat I = \hat D,
\end{equation}
where the following matrix notation  were used
\begin{eqnarray}
\label{eq.246} \hat E = 
\begin{pmatrix}
(\omega- \epsilon_{k})& - V_{k} & - V_{k}\\ 
0&(\omega - E_{0\sigma} - U_{+})&0\\
0&0&(\omega - E_{0\sigma} -U_{-})\\
\end{pmatrix}, \\
\hat I = 
\begin{pmatrix} 
1&0&0\\ 
0&n^{+}_{0-\sigma}&0\\
0&0&n^{-}_{0-\sigma} \\  
\end{pmatrix}, \quad
U_{\alpha} = 
\begin{cases}
U, & \alpha = +  \\ 
0,&  \alpha = -  \\ 
\end{cases} 
\end{eqnarray}
Here $\hat D$ is a higher-order Green function, with the following structure~\cite{kuzam91,kuzem96}
\begin{equation}
\label{eq.247}
 \hat D (\omega) =
\begin{pmatrix} 
0&0&0\cr D_{21}&D_{22}&D_{23}\cr
D_{31}&D_{32}&D_{33}\\
\end{pmatrix}.
\end{equation}
In accordance with the general method of irreducible Green functions, we   define
the matrix irreducible Green function:
\begin{equation}
\label{eq.248} 
\hat D^{(ir)}(\omega) = \hat D - \sum_{\alpha} {
A^{+ \alpha} \choose A^{- \alpha} } (G^{\alpha +}_{\sigma} \quad
G^{\alpha - }_{\sigma} ).
\end{equation}
Here the notation were used:
\begin{eqnarray}
\label{eq.249} A^{++} = \frac {
\langle (f^{\dagger}_{0-\sigma}c_{p-\sigma} +
c^{\dagger}_{p-\sigma}f_{0-\sigma})
(n_{0\sigma} - n_{0-\sigma}) \rangle}{ \langle n_{0-\sigma} \rangle},\\
\label{eq.250} A^{--} = \frac {-
\langle(f^{\dagger}_{0-\sigma}c_{p-\sigma} +
c^{\dagger}_{p-\sigma}f_{0-\sigma})
(1 + n_{0\sigma} - n_{0-\sigma}) \rangle}{ \langle 1 - n_{0-\sigma} \rangle},  \\
A^{-+} = A^{++}, \quad A^{+-} = - A^{--}.  
\end{eqnarray}
The generalized mean-field Green function is defined by
\begin{equation}
\label{eq.251} \hat E \hat F^{0}_{\sigma}(\omega) - \hat I = 0;
\quad G^{0} =\sum_{\alpha \beta} F^{0}_{\alpha \beta}.
\end{equation}
From the last definition we find that
\begin{eqnarray}
\label{eq.252} \langle \langle f_{0\sigma} \vert f^{\dagger}_{0\sigma}\rangle \rangle^{0}_
{\omega} = \frac { \langle n_{0-\sigma} \rangle}{ \omega -E_{0\sigma} - U_{+} -
\sum_{p} V_{p}A^{++}} \bigl ( 1 + \frac { \sum_{p}
V_{p}A^{-+}}{\omega -E_{0\sigma} - U_{-}}
\bigr ) \nonumber \\
+ \frac { 1 - \langle n_{0-\sigma} \rangle}{ \omega -E_{0\sigma} - U_{-} -
\sum_{p} V_{p}A^{--}} \bigl ( 1 + \frac { \sum_{p}
V_{p}A^{+-}}{\omega -E_{0\sigma} - U_{+}}
\bigr ), \\
\label{eq.253}
 \langle \langle c_{k\sigma} \vert c^{\dagger}_{k\sigma}\rangle \rangle^{0}_{\omega} = \bigl (
\omega -\epsilon_{k} -  |V_{k}|^{2}F^{at}(\omega ) \bigr )^{-1},
\end{eqnarray}
where
\begin{equation}
\label{eq.254} F^{at} = \frac { \langle n_{0-\sigma} \rangle}{ \omega
-E_{0\sigma} - U_{+} } + \frac { 1 - \langle n_{0-\sigma} \rangle}{ \omega
-E_{0\sigma} - U_{-} }
\end{equation}
For $V_{p} = 0$, we obtain, from solution (\ref{eq.252}), the
atomic solution $F^{at}$. The conduction electron Green function
(\ref{eq.253}) also gives a correct expression for $V_{k} = 0$.
%
%
%%%%%%%%%%%%%%%%%%%%%%%%%%%%%%%%%%%%%%%%%%%%%%%%%%%%%%%
\section{IGF Method and Interpolation Solution of SIAM}
%%%%%%%%%%%%%%%%%%%%%%%%%%%%%%%%%%%%%%%%%%%%%%%%%%%%%%%
%
%
To show  explicitly the flexibility of the IGF method, we
consider  a more extended  new algebra of operators from which the
relevant matrix Green function should be constructed to make the connection
with the interpolation solution of the Anderson model~\cite{kuzem96}. Our approach
was stimulated by the works by J. Hubbard~\cite{hubb4,hubb5,hubb6}.\\
Let us consider the following equation of motion in the matrix form
\begin{equation}
\label{eq.255} \sum_{p} \hat F(p,k) \hat G_{p\sigma}(\omega) = \hat I + \sum_{p}
V_{p} \hat D_{p}(\omega),
\end{equation}
where $ \hat G$ is the initial $4 \times 4$ matrix Green function and $D$ is the
higher-order Green function:
\begin{equation}
\label{eq.256} \hat G_{\sigma} =
\begin{pmatrix}
G_{11}&G_{12}&G_{13}&G_{14}\cr G_{21}&G_{22}&G_{23}&G_{24}\\
G_{31}&G_{32}&G_{33}&G_{34}\cr G_{41}&G_{42}&G_{43}&G_{44}\\
\end{pmatrix}.
\end{equation}
Here the following notation were used
\begin{eqnarray}
\label{eq.257} G_{11} = \langle \langle c_{k\sigma}|c^{\dagger}_{k\sigma} \rangle \rangle;
\quad G_{12} = \langle \langle c_{k\sigma}|f^{\dagger}_{0\sigma} \rangle \rangle;
\nonumber \\
G_{13} = \langle \langle c_{k\sigma}|f^{\dagger}_{0\sigma}n_{0-\sigma} \rangle \rangle; \quad
G_{14} = \langle \langle c_{k\sigma}|c^{\dagger}_{k\sigma}n_{0-\sigma} \rangle \rangle;
\nonumber \\
G_{21} = \langle \langle f_{0\sigma}|c^{\dagger}_{k\sigma}\rangle \rangle; \quad G_{22} =
\langle \langle f_{0\sigma}|f^{\dagger}_{0\sigma}\rangle \rangle;
\nonumber \\
G_{23} = \langle \langle f_{0\sigma}|f^{\dagger}_{0\sigma}n_{0-\sigma}\rangle \rangle; \quad
G_{24} =
\langle \langle f_{0\sigma}|c^{\dagger}_{k\sigma}n_{0-\sigma}\rangle \rangle; \\
G_{31} = \langle \langle f_{0\sigma}n_{0-\sigma}|c^{\dagger}_{k\sigma} \rangle \rangle; \quad
G_{32} = \langle \langle f_{0\sigma}n_{0-\sigma}|f^{\dagger}_{0\sigma} \rangle \rangle;
\nonumber \\
G_{33} =
\langle \langle f_{0\sigma}n_{0-\sigma}|f^{\dagger}_{0\sigma}n_{0-\sigma}\rangle \rangle;
\quad G_{34} =
\langle \langle f_{0\sigma}n_{0-\sigma}|c^{\dagger}_{k\sigma}n_{0-\sigma}\rangle \rangle;
\nonumber \\
G_{41} = \langle \langle c_{k\sigma}n_{0-\sigma}|c^{\dagger}_{k\sigma}\rangle \rangle; \quad
G_{42} = \langle \langle c_{k\sigma}n_{0-\sigma}|f^{\dagger}_{0\sigma}\rangle \rangle;
\nonumber \\
G_{43} =
\langle \langle c_{k\sigma}n_{0-\sigma}|f^{\dagger}_{0\sigma}n_{0-\sigma}\rangle \rangle;
\quad G_{44} =
\langle \langle c_{k\sigma}n_{0-\sigma}|c^{\dagger}_{k\sigma}n_{0-\sigma}\rangle \rangle.
\nonumber
\end{eqnarray}
We avoid to write down explicitly the relevant 16 Green functions, of which
the matrix Green function $D$ consist,   for the brevity. For our aims, it is
enough  to proceed forth in the following way. \\ The equation
(\ref{eq.255}) results from the first-time differentiation of the
Green function $G$ and is a starting point for the IGF approach. Let us
introduce the irreducible part for the higher-order Green function $D$ in the following way
\begin{equation}
\label{eq.258} \hat D^{(ir)}_{\beta} = \hat D_{\beta} -
\sum_{\alpha}\hat  L^{\beta \alpha} \hat G_{\alpha \beta}; \quad ({\alpha,
\beta}) = (1,2,3,4)
\end{equation}
and define the generalized mean field Green function according to
\begin{equation}
\label{eq.259} \sum_{p}{  \hat F(p,k)} \hat G^{MF}_{p\sigma}(\omega) = \hat I. 
\end{equation} 
Then, we are able to write down explicitly the
Dyson equation (\ref{e78}) and the exact expression for the self-energy
$M$  in the matrix form:
\begin{equation}
\label{eq.260} \hat M_{k\sigma}(\omega) = \hat I^{-1}\sum_{p,q}
V_{p}V_{q}
\begin{pmatrix}
0&0&0&0 \\ 
0&0&0&0 \\ 
0&0&M_{33}&M_{34} \\
0&0&M_{43}&M_{44} \\
\end{pmatrix}
\hat I^{-1}. 
\end {equation} 
Here the matrix $\hat I$ is given by 
\begin{equation}
\hat  I = 
\begin{pmatrix} 
 1&0&0& \langle n_{0-\sigma} \rangle \\
0&1&\langle n_{0-\sigma} \rangle&0 \\ 
0&\langle n_{0-\sigma} \rangle & \langle n_{0-\sigma} \rangle & 0 \\
\langle n_{0-\sigma} \rangle & 0 & 0 & \langle n_{0-\sigma} \rangle \\
\end{pmatrix},
\end{equation}
and the  matrix elements of $M$ are of the form:
\begin{eqnarray}
\label{eq.261} \nonumber
 M_{33} = (\langle \langle A^{(ir)}_{1}(p)|B^{(ir)}_{1}(q)\rangle \rangle)^{(p)},\quad 
M_{34} = (\langle \langle A^{(ir)}_{1}(p)| B^{(ir)}_{2}(k,q)\rangle \rangle)^{(p)}  \\ \nonumber
M_{43} = (\langle \langle A^{(ir)}_{2}(k,p)|B^{(ir)}_{1}(q)\rangle \rangle)^{(p)}, \quad   M_{44} =
(\langle \langle A^{(ir)}_{2}(k,p)|B^{(ir)}_{2}(k,q)\rangle \rangle)^{(p)}. \\
\end{eqnarray}
Here
\begin{eqnarray}
\label{eq.262} A_{1}(p) = (
c^{\dagger}_{p-\sigma}f_{0\sigma}f_{0-\sigma} -
c_{p-\sigma}f^{\dagger}_{0-\sigma}f_{0\sigma} );
\nonumber \\
A_{2}(k,p) = ( c_{k\sigma}f^{\dagger}_{0-\sigma}c_{p-\sigma} -
c_{k\sigma}c^{\dagger}_{p-\sigma}f_{0-\sigma} );
\nonumber \\
B_{1}(p) = (
f^{\dagger}_{0\sigma}c^{\dagger}_{p-\sigma}f_{0-\sigma} -
f^{\dagger}_{0\sigma}f^{\dagger}_{0-\sigma}c_{p-\sigma} ); \\
B_{2}(k,p) = (
c^{\dagger}_{k\sigma}c^{\dagger}_{p-\sigma}f_{0-\sigma} -
c^{\dagger}_{k\sigma}f^{\dagger}_{0-\sigma}c_{p-\sigma} ).
\nonumber
\end{eqnarray}
Since the self-energy $M$ describes the processes of inelastic
scattering of electrons ($c-c$ , $f-f$, and $c-f$ types), its
approximate representation would be defined by the nature of
physical assumptions about this scattering.\\
To get an idea about the functional structure of our generalized mean field solution (\ref{eq.259}),
let us write down the matrix element $G^{MF}_{33}$:
\begin{eqnarray}
\label{eq.263} G^{MF}_{33} =
\langle \langle f_{0\sigma}n_{0-\sigma}|f^{\dagger}_{0\sigma}n_{0-\sigma}\rangle \rangle = \nonumber \\
\frac {\langle n_{0-\sigma} \rangle}{\omega -\epsilon^{MF}_{f} - U -
S^{MF}(\omega) - Y(\omega)} + \nonumber \\
\frac {\langle n_{0-\sigma} \rangle Z(\omega)}{(\omega - \epsilon^{MF}_{f} - U -
S^{MF}(\omega) - Y(\omega))(\omega - E_{0\sigma} - S(\omega))}; \\
\label{eq.264}
Y(\omega) = \frac {UZ(\omega)}{\omega - E_{0\sigma} - S(\omega)}; \\
\label{eq.265} Z(\omega) =
S(\omega)\sum_{p}\frac{V_{P}L^{41}}{\omega -
\epsilon^{MF}_{p}} + 
\sum_{p}\frac{|V_{p}|^{2}L^{42}}{\omega -\epsilon^{MF}_{p}}
+S(\omega)L^{31} + \sum_{p}V_{p}L^{32}.
\end{eqnarray}
Here the coefficients $L^{41}, L^{42}, L^{31}$, and $L^{32}$ are
certain complicated averages (see definition (\ref{eq.258}))
from which the functional of the generalized mean field is build. To clarify the
functional structure of the obtained solution, let us consider our
first equation of motion (\ref{eq.255}), before introducing the
irreducible Green functions (\ref{eq.258}). Let us put  in this
equation  the higher-order Green function $ D = 0$. To distinguish this
simplest equation from the generalized mean field one (\ref{eq.259}), we write it in
the following form
\begin{equation}
\label{eq.266}
 \sum_{p} \hat F(p,k) \hat G^{0}(p,\omega) = \hat I. 
\end{equation}
The corresponding matrix elements   which we are interested in here read
\begin{eqnarray}
\label{eq.267} G^{0}_{22} = \langle \langle f_{0\sigma}|f^{\dagger}_{0\sigma}\rangle \rangle = \\
\frac{1 - \langle n_{0-\sigma} \rangle}{\omega - E_{0\sigma} - S(\omega)} +
\frac{\langle n_{0-\sigma} \rangle}{\omega - E_{0\sigma} -S(\omega) - U},  \nonumber \\
\label{eq.268} G^{0}_{33} =
\langle \langle f_{0\sigma}n_{0-\sigma}|f^{\dagger}_{0\sigma}n_{0-\sigma}\rangle \rangle =
\frac{\langle n_{0-\sigma} \rangle}{\omega - E_{0\sigma} - S(\omega) - U},\\
\label{eq.269} G^{0}_{32} =
\langle \langle f_{0\sigma}n_{0-\sigma}|f^{\dagger}_{0\sigma}\rangle \rangle = G^{0}_{33}.
\end{eqnarray}
The conclusion is rather evident. The simplest interpolation
solution  follows from our matrix  Green function (\ref{eq.256}) in the lowest
order in $V$, even before introduction of generalized mean field corrections, not
speaking about   the self-energy corrections. The two Green functions
$G^{0}_{32}$ and $G^{0}_{33}$  are equal only in the lowest order
in $V$. It is quite clear  that our  full  solution (\ref{e79}) that
includes the self-energy corrections   is much more
richer.\\
It is worthwhile to stress that our $4\times4$ matrix generalized mean field Green function
(\ref{eq.256}) gives only approximate description of   suitable
mean fields. If we   consider more extended algebra of relevant operators, we   get the
more correct structure of the relevant generalized mean field. 
%
%%%%%%%%%%%%%%%%%%%%%%%%%%%%%%%%%%%%%%%%%%%  
\section{Quasiparticle Dynamics   of SIAM}
%%%%%%%%%%%%%%%%%%%%%%%%%%%%%%%%%%%%%%%%%%%% 
%
%
To demonstrate   more clearly the advantages of the irreducible Green functions method for SIAM,
it is worthwhile to emphasize a few important points about the
approach   based on the equations-of-motion for the Green functions. To give
a more instructive discussion, let us consider the single-particle
Green function of localized electrons $ G_{\sigma} = \langle \langle f_{0\sigma}\vert
f^{\dagger}_{0\sigma}\rangle \rangle $. The simplest approximate
"interpolation" solution  of SIAM is of the form:
\begin{eqnarray}
\label{eq.270} G_{\sigma}(\omega) = \frac{1}{\omega - E_{0\sigma}
- S(\omega)} +
 \frac{U\langle n_{0-\sigma} \rangle}{(\omega - E_{0\sigma} - S(\omega) - U)(\omega -
E_{0\sigma} - S(\omega))} = \nonumber \\ \frac{1 -
\langle n_{0-\sigma} \rangle}{\omega - E_{0\sigma} - S(\omega) } +
\frac{\langle n_{0-\sigma} \rangle}{\omega - E_{0\sigma} - S(\omega) - U}.
\end{eqnarray}
The values of $n_{\sigma}$  are determined through the
self-consistency equation
\begin{equation}
\label{eq.271} n_{\sigma} = \langle n_{0\sigma} \rangle = -\frac {1}{\pi} \int
dE \, n(E)\textrm{Im} \, G_{\sigma}(E,n_{\sigma}).
\end{equation}
The \emph{atomic-like}  interpolation solution (\ref{eq.270}) reproduces
correctly the two  limits:
\begin{eqnarray}
\label{eq.272} G_{\sigma}(\omega) = \frac{1 -
\langle n_{0-\sigma} \rangle}{\omega - E_{0\sigma}} +
\frac{\langle n_{0-\sigma} \rangle}{\omega - E_{0\sigma} - U},\quad \textrm{for} \quad V = 0,   \\ 
\label{eq.272a}
G_{\sigma}(\omega) = \frac{1}{\omega -
E_{0\sigma} - S(\omega)},\quad \textrm{for} \quad U = 0,
\end{eqnarray}
where
\begin{equation}
\label{eq.231a} S(\omega) = \sum_{k} \frac { |V_{k}|^{2}}{\omega - \epsilon_{k}}.
\end{equation}
The important point about equations (\ref{eq.272}),(\ref{eq.272a})  is that any
approximate solution of SIAM should be consistent with it. Let us
remind how to get solution (\ref{eq.272}). It follows from the
system of equations for small-$V$ limit:
\begin{eqnarray}
\label{eq.273} (\omega - E_{0\sigma} - S(\omega))
\langle \langle f_{0\sigma}|f^{\dagger}_{0\sigma}\rangle \rangle_{\omega} = 1 +
U\langle \langle f_{0\sigma}n_{0-\sigma}|f^{\dagger}_{0\sigma}\rangle \rangle_{\omega},  \nonumber \\
(\omega - E_{0\sigma} -
U)\langle \langle f_{0\sigma}n_{0-\sigma}|f^{\dagger}_{0\sigma}\rangle \rangle _{\omega}
\approx \nonumber \\ \langle n_{0-\sigma} \rangle + \sum_{k}
V_{k}\langle \langle c_{k\sigma}n_{0\sigma}|f^{\dagger}_{0\sigma}\rangle \rangle_{\omega}, \\
\label{eq.273a}
(\omega - \epsilon_{k})\langle \langle c_{k\sigma}n_{0-\sigma}|f^{\dagger}_{0\sigma}\rangle \rangle_{
\omega} =  
V_{k}\langle \langle f_{0\sigma}n_{0-\sigma}|f^{\dagger}_{0\sigma}\rangle \rangle_{\omega}.
\end{eqnarray}
Note that the equations (\ref{eq.273}) and (\ref{eq.273a})  are approximate; they include two
more terms.   \\ We now proceed further.  The starting point is the system of equations:
\begin{eqnarray}
\label{eq.274} (\omega - E_{0\sigma} -
S(\omega))\langle \langle f_{0\sigma}|f^{\dagger}_{0\sigma}\rangle \rangle =
1 + U\langle \langle f_{0\sigma}n_{0-\sigma}|f^{\dagger}_{0\sigma}\rangle \rangle, \\
(\omega - E_{0\sigma}
-U)\langle \langle f_{0\sigma}n_{0-\sigma}|f^{\dagger}_{0\sigma}\rangle \rangle =
\langle n_{0-\sigma} \rangle +\sum_{k}V_{k} \Bigl
(\langle \langle c_{k\sigma}n_{0-\sigma}|f^{\dagger}_{0\sigma}\rangle \rangle - \nonumber
\\
\label{eq.275}
\langle \langle c_{k-\sigma}f^{\dagger}_{0-\sigma}f_{0\sigma}|f^{\dagger}_{0\sigma}\rangle \rangle
+
\langle \langle c^{\dagger}_{k-\sigma}f_{0\sigma}f_{0-\sigma}|f^{\dagger}_{0\sigma}\rangle \rangle
\Bigr ).
\end{eqnarray}
Using a relatively simple decoupling procedure for a higher-order
equation of motion,  a qualitatively correct low-temperature
spectral intensity can be calculated. The final expression for $G$
for finite $U$ is of the form
\begin{eqnarray}
\label{eq.276} \langle \langle f_{0\sigma}|f^{\dagger}_{0\sigma}\rangle \rangle =
\frac{1}{\omega - E_{0\sigma} -
S(\omega) +US_{1}(\omega)} + \nonumber \\
\frac{U\langle n_{0-\sigma} \rangle +UF_{1}(\omega)}{K(\omega)(\omega -
E_{0\sigma} - S(\omega) + US_{1}(\omega))},
\end{eqnarray}
where $F_{1}$, $S_{1}$, and $K$ are certain complicated
expressions. We   write down explicitly the infinite $U$
approximate Green function:
\begin{equation}
\label{eq.277} \langle \langle f_{0\sigma}|f^{\dagger}_{0\sigma}\rangle \rangle  \cong  \frac{1 -
\langle n_{0-\sigma} \rangle - F_{\sigma}( \omega)}{\omega - E_{0\sigma} -
S(\omega) - Z^{1}_{\sigma}(\omega)}.
\end{equation}
The following notation  were used:
\begin{eqnarray}
\label{eq.278} F_{\sigma} = V\sum_{k}
\frac{\langle f^{\dagger}_{0-\sigma}c_{k-\sigma} \rangle}{\omega - \epsilon_{k}},\\
\label{eq.279} Z^{1}_{\sigma} = V^{2}\sum_{q,k}
\frac{\langle c^{\dagger}_{q-\sigma}c_{k-\sigma} \rangle}{\omega -
\epsilon_{k}} - S(\omega)V \sum_{k}
\frac{\langle f^{\dagger}_{0-\sigma}c_{k-\sigma} \rangle}{\omega - \epsilon_{k}}.
\end{eqnarray} 
We putted here $V_{k} \simeq V$ for brevity.
The functional structure of the single-particle Green function
(\ref{eq.276}) is quite transparent. The expression in the
numerator of (\ref{eq.276}) plays the role of  an effective  \emph{dynamical mean field},   
proportional to $\langle f^{\dagger}_{0-\sigma}c_{k-\sigma} \rangle$.
In the denominator, instead of bare shift $S(\omega)$
(\ref{eq.231})  we have an  \emph{effective shift}  $S^{1} = S(\omega) +
Z^{1}_{\sigma}(\omega)$. The  choice of the specific procedure of
decoupling for the higher-order equation of motion specifies the
selected  \emph{generalized mean fields}    and  \emph{effective shifts}.
%
%
%
%%%%%%%%%%%%%%%%%%%%%%%%%%%%%%%%%%%%%%%%%%%%%%%%%%%%%%
\section{Complex Expansion for a Propagator}
%%%%%%%%%%%%%%%%%%%%%%%%%%%%%%%%%%%%%%%%%%%%%%%%%%%%%%
%
We now proceed with analytic  many-body consideration.  One can
attempt to consider a suitable solution for the SIAM starting
from the following exact relation   derived in
paper~\cite{kuzem96}:
\begin{eqnarray} \label{eq.289}
\langle \langle f_{0\sigma}|f^{\dagger}_{0\sigma}\rangle \rangle = g^{0} + g^{0}Pg^{0}, \\
\label{eq.290}
g^{0} = (\omega - E_{0\sigma} - S(\omega))^{-1},\\
\label{eq.291} P = U\langle n_{0-\sigma} \rangle +
U^{2}\langle \langle f_{0\sigma}n_{0-\sigma}|f^{\dagger}_{0\sigma}n_{0-\sigma}\rangle \rangle.
\end{eqnarray}
The advantage of the equation (\ref{eq.289}) is that it is a
pure  \textbf{identity} and does not includes any approximation. If we
insert our generalized mean field solution (\ref{eq.277}) into (\ref{eq.289}), we
get an essentially new dynamic solution of SIAM
constructed on the basis of the complex (combined) expansion of
the propagator in both $U$ and $V$ parameters and   reproducing
  exact solutions of SIAM for $V = 0$ and $U = 0 $. It
generalizes (even on the mean-field level) the known approximate solutions of
the Anderson model.\\
Having emphasized the importance of the role of   equation
(\ref{eq.289}) , let us see now what is the best possible fit for
higher-order Green function in (\ref{eq.291}). We   consider the equation of motion for it:
\begin{eqnarray}
\label{eq.292}
 (\omega - E_{0\sigma} -
U)\langle \langle f_{0\sigma}n_{0-\sigma}|f^{\dagger}_{0\sigma}n_{0-\sigma}\rangle \rangle =
\langle n_{0-\sigma} \rangle +  \\
\sum_{k}V_{k}(\langle \langle c_{k\sigma}n_{0-\sigma}|f^{\dagger}_{0\sigma}n_{0-\sigma}\rangle \rangle
+ \nonumber
\\
\langle \langle c^{\dagger}_{k-\sigma} f_{0\sigma}f_{0-\sigma}|f^{\dagger}_{0\sigma} n_{0-\sigma}\rangle \rangle
-
\langle \langle c_{k-\sigma}f^{\dagger}_{0-\sigma}f_{0\sigma}|f^{\dagger}_{0\sigma} n_{0-\sigma}\rangle \rangle).
\nonumber
\end{eqnarray}
We may think of it as defining   new kinds of elastic and
inelastic scattering processes that contribute to the formation
of   generalized mean fields and self-energy (damping)
corrections. The construction of   suitable mean fields can be
quite nontrivial, and to describe these contributions
self-consistently, let us consider the equations of motion for
higher-order Green functions in the r.h.s. of (\ref{eq.292})
\begin{eqnarray}
\label{eq.293} (\omega -
\epsilon_{k})\langle \langle c_{k\sigma}n_{0-\sigma}|f^{\dagger}_{0\sigma}n_{0-\sigma}\rangle \rangle
=  \nonumber \\
V \langle \langle f_{0\sigma}n_{0-\sigma}|f^{\dagger}_{0\sigma}n_{0-\sigma}\rangle \rangle +  \\
\sum_{p} V
(\langle \langle c_{k\sigma}f^{\dagger}_{0-\sigma}c_{p-\sigma}|f^{\dagger}_{0\sigma}n_{0-\sigma}\rangle \rangle
-
\langle \langle c_{k\sigma}c^{\dagger}_{p-\sigma}f_{0-\sigma}|f^{\dagger}_{0\sigma}n_{0-\sigma}\rangle \rangle),
\nonumber
\end{eqnarray}
\begin{eqnarray}
\label{eq.294} (\omega - \epsilon_{k} -E_{0\sigma} + E
_{0-\sigma})\langle \langle c_{k-\sigma}f^{\dagger}_{0-\sigma}f_{0\sigma}|f^{\dagger}_{0\sigma}n_{0-\sigma}\rangle \rangle
\nonumber \\
= - \langle f^{\dagger}_{0-\sigma}c_{k-\sigma}n_{0\sigma} \rangle -  \\
V \langle \langle f_{0\sigma}n_{0-\sigma}|f^{\dagger}_{0\sigma}n_{0-\sigma}\rangle \rangle + \nonumber \\
\sum_{p}V(\langle \langle c_{k-\sigma}f^{\dagger}_{0-\sigma}c_{p\sigma}|f^{\dagger}_{0\sigma}n_{0-\sigma}\rangle \rangle
-
\langle \langle c_{k-\sigma}c^{\dagger}_{p-\sigma}f_{0\sigma}|f^{\dagger}_{0\sigma}n_{0-\sigma}\rangle \rangle),
\nonumber
\end{eqnarray}
\begin{eqnarray}
\label{eq.295} (\omega + \epsilon_{k} - E_{0\sigma} - E
_{0-\sigma} -
U)\langle \langle c^{\dagger}_{k-\sigma}f_{0\sigma}f_{0-\sigma}|f^{\dagger}_{0\sigma}n_{0-\sigma}\rangle \rangle
\\ = - \langle c^{\dagger}_{k-\sigma}f_{0\sigma}f^{\dagger}_{0\sigma}f_{0-\sigma} \rangle +
\nonumber
\\
V\langle \langle f_{0\sigma}n_{0-\sigma}|f^{\dagger}_{0\sigma}n_{0-\sigma}\rangle \rangle +
\nonumber
\\
\sum_{p}V(\langle \langle c^{\dagger}_{k-\sigma}c_{p\sigma}f_{0-\sigma}|f^{\dagger}_{0\sigma}n_{0-\sigma}\rangle \rangle
+
\langle \langle c^{\dagger}_{k-\sigma}f_{0\sigma}c_{p-\sigma}|f^{\dagger}_{0\sigma}n_{0-\sigma}\rangle \rangle).
\nonumber
\end{eqnarray}
Now let us see how to proceed further to get a suitable
functional structure of the relevant solution. The intrinsic
nature of the system of the equations of motion (\ref{eq.293}) -
(\ref{eq.295}) suggests to consider the following approximations:
\begin{align}
\label{eq.296} (\omega -
\epsilon_{k})\langle \langle c_{k\sigma}n_{0-\sigma}|f^{\dagger}_{0\sigma}n_{0-\sigma}\rangle \rangle
\approx
V\langle \langle f_{0\sigma}n_{0-\sigma}|f^{\dagger}_{0\sigma}n_{0-\sigma}\rangle \rangle,\\
\label{eq.297} (\omega - \epsilon_{k} - E_{0\sigma} +
E_{0-\sigma})\langle \langle c_{k-\sigma}f^{\dagger}_{0-\sigma}f_{0\sigma}|
f^{\dagger}_{0\sigma}n_{0-\sigma}\rangle \rangle \approx
- \langle f^{\dagger}_{0-\sigma}c_{k-\sigma}n_{0\sigma} \rangle \nonumber \\ -
V(\langle \langle f_{0\sigma}n_{0-\sigma}|f^{\dagger}_{0\sigma}n_{0-\sigma}\rangle \rangle -
\langle \langle c_{k-\sigma}c^{\dagger}_{k-\sigma}f_{0\sigma}|f^{\dagger}_{0\sigma}n_{0-\sigma}\rangle \rangle),\\
\label{eq.298} (\omega + \epsilon_{k} - E_{0\sigma} - E_{0-\sigma}
-
U)\langle \langle c^{\dagger}_{k-\sigma}f_{0\sigma}f_{0-\sigma}|f^{\dagger}_{0\sigma}n_{0-\sigma}\rangle \rangle
\approx
- \langle c^{\dagger}_{k-\sigma}f_{0\sigma}f^{\dagger}_{0\sigma}f_{0-\sigma} \rangle
+\nonumber \\
V(\langle \langle f_{0\sigma}n_{0-\sigma}|f^{\dagger}_{0\sigma}n_{0-\sigma}\rangle \rangle +
\langle \langle c^{\dagger}_{k-\sigma}f_{0\sigma}c_{k-\sigma}|f^{\dagger}_{0\sigma}n_{0-\sigma}\rangle \rangle).
\end{align}
It is transparent that the construction of   approximations
(\ref{eq.296}) - (\ref{eq.298}) is related with the small-$V$
expansion and is not unique, but very natural. As a result, we
find the explicit expression for   Green function in (\ref{eq.291})
\begin{equation}
\label{eq.299}
\langle \langle f_{0\sigma}n_{0-\sigma}|f^{\dagger}_{0\sigma}n_{0-\sigma}\rangle \rangle
\approx \frac{\langle n_{0-\sigma} \rangle - F^{1}_{\sigma}(\omega)}{\omega -
E_{0\sigma} - U - S_{1}(\omega)}.
\end{equation}
Here the following notation were used:
\begin{eqnarray}
\label{eq.300} S_{1}(\omega) = S(\omega) \\ +
\sum_{k}|V|^{2}(\frac{1}{\omega - \epsilon_{k} - E_{0\sigma} +
E_{0-\sigma}} + \frac{1}{\omega + \epsilon_{k}
- E_{0\sigma} - E_{0-\sigma} - U}), \nonumber \\
\label{eq.301}
F^{1}_{\sigma} = \sum_{k}(VF_{2} + V^{2}F_{3}),\\
\label{eq.302} F_{2} =
\frac{\langle c^{\dagger}_{k-\sigma}f_{0\sigma}f^{\dagger}_{0\sigma}f_{0-\sigma} \rangle}{\omega
+ \epsilon_{k} - E_{0\sigma} - E_{0-\sigma} - U } +
\frac{\langle f^{\dagger}_{0-\sigma}c_{k-\sigma}n_{0\sigma} \rangle}{\omega -
\epsilon_{k} -
E_{0\sigma} + E_{0-\sigma}},\\
\label{eq.303}
F_{3} =  
\frac{\langle \langle c_{k-\sigma}c^{\dagger}_{k-\sigma}f_{0\sigma}|f^{\dagger}_{0\sigma}n_{0-\sigma}\rangle \rangle}{
\omega - \epsilon_{k} - E_{0\sigma} + E_{0-\sigma}} +
\frac{\langle \langle c^{\dagger}_{k-\sigma}f_{0\sigma}c_{k-\sigma}|f^{\dagger}_{0\sigma}n_{0-\sigma}\rangle \rangle}{
\omega + \epsilon_{k} - E_{0\sigma} - E_{0-\sigma} - U}.  
\end{eqnarray}
Now one can substitute the Green function in (\ref{eq.291}) by the expression
(\ref{eq.299}). This   gives  a new approximate dynamic solution
of SIAM where the complex expansion  both in $U$ and $V$ was
incorporated. The important observation is that this new solution
satisfies  both the limits (\ref{eq.272}).  For example, if we
wish to get a lowest order approximation up to $U^{2}$ and
$V^{2}$, it is very easy to notice that for $V = 0$:
\begin{eqnarray}
\label{eq.304}
\langle \langle f_{0\sigma}c^{\dagger}_{k-\sigma}c_{k-\sigma}|f^{\dagger}_{0\sigma}n_{0-\sigma}\rangle \rangle
\approx
\frac{\langle c^{\dagger}_{k-\sigma}c_{k-\sigma} \rangle \langle n_{0-\sigma} \rangle}{\omega -
E_{0\sigma} - U},  \\
\langle \langle c_{k-\sigma}c^{\dagger}_{k-\sigma}f_{0\sigma}|f^{\dagger}_{0\sigma}n_{0-\sigma}\rangle \rangle
\approx
\frac{\langle c_{k-\sigma}c^{\dagger}_{k-\sigma}\rangle \langle n_{0-\sigma} \rangle}{\omega -
E_{0\sigma} - U}.
\end{eqnarray}
This results in the possibility to find explicitly all necessary
quantities and, thus, to solve the problem in a self-consistent
way. The main results of our IGF study is the exact Dyson equation 
for the full  matrix Green function  and a new derivation of 
the \emph{generalized mean field} Green functions.
The approximate explicit calculations of   inelastic self-energy
corrections are quite straightforward but tedious and too
extended for their description. Here we want to emphasize an
essentially new point of view on the derivation of the
generalized mean fields for SIAM when we are interested in the
interpolation finite temperature solution for the single-particle
propagator.  Our final solutions have the correct functional
structure  and differ essentially from previous solutions.\\
In summary, we   presented here a consistent many-body approach
to analytic  dynamic solution of SIAM  at finite temperatures and
for a broad interval of the values of the model parameters.  We
used the exact result (\ref{eq.289}) to connect the
single-particle Green function with   higher-order Green function to obtain a  complex
combined expansion in terms of $U$ and $V$ for the propagator.  
We   reformulated also the problem of searches for an
appropriate  many-body dynamic solution for SIAM in a way that
provides us with an effective and workable scheme for
constructing of advanced analytic  approximate solutions for the
single-particle Green functions on the level of the higher-order Green functions in a
rather systematic   self-consistent way. This procedure has the
advantage that it systematically uses the principle of
interpolation solution within the equation-of-motion approach for
  Green functions. The leading principle, which we   used here was to
look more carefully for the intrinsic functional structure of the
required relevant solution and then to formulate approximations
for the higher-order Green functions in accordance with this structure. \\
Of course, there are important
criteria to be met (mainly numerically) , such as the question
left open, whether the present approximation satisfies the Friedel
sum rule (this question was left open in many other approximate solutions). 
A quantitative numerical comparison of self-consistent
results e.g.  the width and shape of the Kondo resonance in
the near-integer regime of the SIAM  would be crucial too. In the
present consideration, we   concentrated on the problem of
correct functional structure of the single-particle Green function itself. 
%
%
%
%
%%%%%%%%%%%%%%%%%%%%%%%%%%%%%%%%%%%%%%%%%%%%%%%%%%%%%% 
\section{The Improved Interpolative Treatment of SIAM}
%%%%%%%%%%%%%%%%%%%%%%%%%%%%%%%%%%%%%%%%%%%%%%%%%%%%%% 
%
%
%
%
For better understanding of the correct functional structure of the single-particle Green function 
the development of improved and reliable approximation schemes is still
justified and necessary, and an effective interpolating approximations are desirable.
The present section is devoted to the development of an improved
interpolating  approximation~\cite{ckw96,kuzem96} for the dynamical properties of the SIAM.
We will show that a self-consistent  approximation can be
formulated which reproduces all relevant exactly solvable limits of the
model and interpolates between the strong- and the weak-coupling limit. This approach is complementary
to the one described above.\\
We start by considering the equations of motion for the Fourier
transformed Green  function,
\begin{eqnarray}
G_{\sigma}(\omega) = \langle \langle f_{\sigma}|f_{\sigma}^\dagger
\rangle \rangle_{\omega} = -i \int_0^{\infty} dt \exp(i \omega t)
\langle [f_{0\sigma}(t),f^{\dagger}_{0\sigma}]_{+} \rangle:
\end{eqnarray}
\begin{eqnarray}
(\omega - E_{\sigma} - S(\omega) ) \langle \langle f_{\sigma}| f^{\dagger}_{\sigma}
\rangle \rangle_{\omega} = 1 + U \langle \langle f_{\sigma} n_{-\sigma}|
f^{\dagger}_{\sigma} \rangle \rangle_{\omega} = 1 + \Sigma_ {\sigma}(\omega) \langle \langle
f_{\sigma}|f^{\dagger}_{\sigma} \rangle \rangle_{\omega}. \quad
\end{eqnarray}
Here the the quantity   $\Sigma_ {\sigma}(\omega)$ may be conditionally interpreted as the one-particle self-energy
and  
\begin{equation}
 S(\omega) = \sum_{k}\frac{|V|^{2}}{\omega - \epsilon_{k}}. 
\end{equation}
We want to develop an  \emph{interpolating}  solution for the SIAM, i.e. a
solution which is applicable in both, the weak-coupling limit (and thus
the exactly solvable band limit) and the strong-coupling limit (and thus
the atomic limit).
As it was shown earlier, the simplest approximative \emph{interpolating} solution has the form:
\begin{eqnarray}
G_{\sigma}(\omega) = \frac{1 - \langle n_{-\sigma} \rangle}{\omega - E_{\sigma} -
S(\omega) } + \frac{\langle n_{-\sigma} \rangle}{\omega - E_{\sigma} - S(\omega) - U}.
\label{HUBIII}\end{eqnarray}
Here $\langle n_{-\sigma} \rangle$ denotes the occupation number of $f-$electrons with
spin $\sigma$. This is just the analogue of the \emph{Hubbard III} approximation
\cite{hubb3} for the SIAM. As for the Hubbard model, however, Fermi liquid  
properties and the Friedel sum rule, which hold for the SIAM  at
least order by order within the $U$-perturbation theory, are violated within
this simple approximation.\\
An approximation, which automatically fulfills Fermi liquid properties and sum rules,
is provided by the self-consistent second order $U$-perturbation treatment (SOPT) and is
given by:
\begin{eqnarray}
\Sigma_{\sigma}(i \omega_n) = U \langle n_{-\sigma} \rangle -
\left(\frac{U}{\beta}\right)^2
\sum_{\omega_1, \nu}
G_{\sigma}(i\omega_n + i\nu) G_{-\sigma}(i\omega_1 - i\nu)
G_{-\sigma}(i\omega_1).
\label{SIGMA2}
\end{eqnarray}
Here $\omega_{1}(\nu)$ denote odd (even) Matsubara frequencies and
$\beta = 1/k_B T$. One of our goals is to find some way to incorporate this
SOPT into an interpolating dynamical solution of the SIAM. This means that
the approximation for the self-energy shall be correct up to order $U^2$
perturbationally around the band limit $U=0$ and also the atomic limit $V = 0$
shall be fulfilled.  This is the case for the SOPT around the Hartree-Fock
solution, but only for the symmetric SIAM. For the general
situation (position of the Fermi level relative to $E_{\sigma}$ and
$E_{\sigma} + U$) a heuristic semi-empirical approach  only for constructing such
an approximation has been discussed in literature
Here  our intention is to take into account the self-consistent-SOPT. Furthermore,  the approximation shall
not only fulfill the atomic limit $ V = 0$, but it shall be correct up to order
$V^2$ in a strong-coupling expansion around the atomic limit.\\
The self-consistent inclusion of contributions in second (and
fourth) order perturbation theory around the atomic limit is, in
particular, important to properly account for the Kondo effect within
the SIAM (Kondo temperature scale) and to reproduce the correct
antiferromagnetic behavior   in the strong-coupling limit of the Hubbard model.
Especially the calculation of some magnetic properties  for the
Hubbard model   and the well known Kondo effect for the SIAM shows the
importance of second (and fourth) order perturbation theory around the
atomic limit.\\
It was already mentioned that during the last decades several different refined many-body techniques
have been applied to the SIAM, and many of
these approaches are strong-coupling treatments around the atomic limit
and can be classified as being correct up to a certain power in the
hybridization $V$. When applied to the calculation of static properties
many of these treatments,   give reasonable results.
But for the many-body dynamics the results of most of these approximations are not fully satisfactory, in
particular as Fermi liquid properties and sum rules are violated.
Furthermore, when applied to the finite-$U$ SIAM none of these
approximation schemes  reproduce the SOPT, i.e. these approaches are not correct in the weak-coupling limit
up to order $U^2$.\\
To construct the interpolating approximation~\cite{kuzam91,ckw96,kuzem96,kpb93} for the SIAM fulfilling all
desired properties mentioned above we start from the equation  of motion for the higher
order Green function $\langle \langle f_{\sigma}n_{-\sigma}|f_{\sigma}^\dagger\rangle \rangle_{\omega}$:
\begin{eqnarray}
(\omega - E_{\sigma} - S(\omega) - U)
\langle \langle f_{\sigma}n_{-\sigma}|f_{\sigma}^\dagger\rangle \rangle_{\omega} = \langle n_{-\sigma} \rangle -
U \langle \langle f_{\sigma} f_{-\sigma} f_{-\sigma}^\dagger|  f_{\sigma}^\dagger
n_{-\sigma} \rangle \rangle_{\omega}.
\nonumber\end{eqnarray}
With 
\begin{equation}
[G^{(0)}_{\sigma}(\omega)]^{-1} = \omega - E_{\sigma} - S(\omega) 
\end{equation}
and the self-consistent summation 
\begin{equation}
 [G^{(0)}_{\sigma}(\omega)]^{-1} G_{\sigma}(\omega) = 1 + \Sigma_{\sigma}(\omega) G_{\sigma}(\omega), 
\end{equation}
we derive from this equations of motion the following exact relation
\begin{eqnarray}
\Sigma_{\sigma}(\omega) = \frac{U \langle n_{-\sigma} \rangle + U^2 {\cal Z}
(\omega) \frac{G_{\sigma}(\omega)} {1 + \Sigma_{\sigma}(\omega) G_{\sigma}(\omega)}}
{1 - (U - \Sigma_{\sigma}(\omega))G_{\sigma}(\omega)}.
\label{EQQ1}
\end{eqnarray}
Here  the definition 
\begin{equation}
  \langle \langle f_{\sigma} f_{-\sigma} f_{-\sigma}^\dagger|
f_{\sigma}^\dagger n_{-\sigma} \rangle \rangle_{\omega} = - {\cal Z}(\omega) \frac{G_{\sigma}(\omega)} {1 +
\Sigma_{\sigma}(\omega) G_{\sigma}(\omega)}  
\end{equation}
was introduced.\\
Applying the equations of motion to the higher-order Green function
\begin{equation}
\langle \langle f_{\sigma} f_{-\sigma} f_{-\sigma}^\dagger|  f_{\sigma}^\dagger
n_{-\sigma} \rangle \rangle_{\omega},  
\end{equation}
one obtains for the function ${\cal Z}(z)$ the exact equation
\begin{eqnarray}
{\cal Z}(\omega) =
V  \sum_{\bf k} \left\{ G_{1\sigma}(\underline{\bf k}) -
G_{2\sigma}(\underline{\bf k}) + \frac{V }{\omega - \epsilon_{\bf k}}
\left[G_{3\sigma}(\underline{\bf k}) - G_{4\sigma}(\underline{\bf k})
\right] \right\},
\label{EQQ}
\end{eqnarray}
with $\underline{\bf k} = ({\bf k}, \omega)$ and
\begin{eqnarray}
G_{1\sigma}(\underline{\bf k}) = \langle \langle f_{\sigma}f^{\dagger}_{-\sigma}c_{{\bf k}-\sigma}|f^{\dagger}_{\sigma}n_{-\sigma} \rangle \rangle_{\omega},\\
G_{2\sigma}(\underline{\bf k}) = \langle \langle f_{\sigma} c^{\dagger}_{{\bf k}-\sigma}f_{-\sigma}|f^{\dagger}_{\sigma} n_{-\sigma} \rangle \rangle_{\omega},\\
G_{3\sigma}(\underline{\bf k}) = \sum_{\bf q} \langle \langle c_{\bf k\sigma} f_{-\sigma}c^{\dagger}_{\bf q-\sigma}|f^{\dagger}_{\sigma} n_{-\sigma} \rangle \rangle_{\omega},\\
G_{4\sigma}(\underline{\bf k}) = \sum_{\bf q} \langle \langle c_{\bf k\sigma} c_{\bf q-\sigma}f^{\dagger}_{-\sigma}|f^{\dagger}_{\sigma} n_{-\sigma} \rangle \rangle_{\omega}. 
\end{eqnarray}
Self-consistency in the perturbation theory defines the Green function:
\begin{equation}
  [G^{(0)}_{\sigma}(\omega)]^{-1} G_{\sigma}(\omega) = 1 + \Sigma_{\sigma}(\omega)
G_{\sigma}(\omega), 
\end{equation}
and leads to an infinite order resummation resulting in a self-consistent
approximation.\\
In general, there are several possibilities to incorporate
self-consistency, but most of these possibilities lead once more to an approximation being
exact up to order $V^2$ but not reproducing the weak-coupling limit.
To be exact up to order $V^2$ it is justified to replace the higher order Green functions on the right
hand side of Eq. (\ref{EQQ}) by their lowest order contributions, which are
given by
\begin{eqnarray}
G_{1\sigma}(\underline{\bf k}) &=&
\frac{V }{\epsilon_{\bf k} - E_{-\sigma} - U}
\biggl[
\frac{\langle n_{-\sigma} \rangle[f_{\bf k} - f(E_{-\sigma} + U)] +
\langle n_{-\sigma} \rangle[1-f_{\bf k}]}{\omega - \epsilon_{\bf k} - E_{\sigma}
+ E_{-\sigma}}
- \frac{\langle n_{-\sigma} \rangle[1 - f_{\bf k}]}{\omega -
E_{\sigma} - U} \biggr] + O(V^3),
\nonumber\\
G_{2\sigma}(\underline{\bf k}) &=&
\frac{V }{\epsilon_{\bf k} - E_{-\sigma}}
\biggl[\frac{(1 - \langle n_{-\sigma} \rangle)[f_{\bf k} - f(E_{-\sigma})]
+ [1-f_{\bf k}]
\langle n_{-\sigma} \rangle}{\omega + \epsilon_{\bf k} - E_{\sigma} - E_{-\sigma} - U}
- \frac{\langle n_{-\sigma} \rangle[1 - f_{\bf k}]}{\omega -
E_{\sigma} - U} \biggr] + O(V^3),
\nonumber\\
G_{3\sigma}(\underline{\bf k}) &=& O(V^2),\mbox{ }
G_{4\sigma}(\underline{\bf k}) = O(V^2),
%\nonumber
\label{EQV}
\end{eqnarray}
leading to a finite order $V^2$ perturbation expansion of the self-energy
(\ref{EQQ1}).
Here
$f(E) = \{\exp[(E - \mu)/k_BT]+1\}^{-1}$ is the Fermi
function, $\mu$ the chemical potential and $f_{\bf k} = f(\epsilon_{\bf k})$.\\
For the higher order Green functions
$G_{i\sigma}(\underline{\bf k})$ ($i=1,\ldots,4$)
one can find an approximation which
reproduces the exact relations (\ref{EQV}) in lowest order in $V$ and is
simultaneously exact in lowest order in $U$ (when Wick's theorem is
applicable). One possibility for such an approximation is given by:
\begin{eqnarray}
G_{1\sigma}(\underline{\bf k}) &=& \frac{-\beta^{-2}}
{\langle n_{-\sigma} \rangle \langle n_{\sigma}n_{-\sigma} \rangle}
\sum_{\omega_1, \nu} \langle \langle f_{\sigma}|
n_{-\sigma}f^{\dagger}_{\sigma} \rangle \rangle_{i\omega_n + i \nu}
 \langle \langle c_{{\bf k}-\sigma}|
n_{\sigma}f^{\dagger}_{-\sigma} \rangle \rangle_{i\omega_1 - i \nu}
\nonumber\\
 &\times&  \langle \langle f_{-\sigma}n_{\sigma}|
f^{\dagger}_{-\sigma} \rangle \rangle_{i\omega_1}
+ \frac{\langle f^{\dagger}_{-\sigma}c_{{\bf k}-\sigma}n_{\sigma} \rangle}
{\langle n_{-\sigma} \rangle}
\langle \langle f_{\sigma}n_{-\sigma}|
f^{\dagger}_{\sigma} \rangle \rangle_{i \omega_n}
\label{APPR1}
\\
G_{2\sigma}(\underline{\bf k}) &=&
\frac{-\beta^{-2}}{(1-\langle n_{-\sigma} \rangle) \langle n_{\sigma}n_{-\sigma} \rangle}
\sum_{\omega_1, \nu} \biggl(\frac{\langle n_{-\sigma} \rangle}{\langle n_{-\sigma} \rangle -
\langle n_{\sigma}n_{-\sigma} \rangle}
\langle \langle f_{\sigma}|
f_{-\sigma}f^{\dagger}_{-\sigma}f^{\dagger}_{\sigma}
\rangle \rangle_{i\omega_n+i\nu}
\nonumber\\
 &\times&  \langle \langle f_{-\sigma}|
   n_{\sigma}f^{\dagger}_{-\sigma}
   \rangle \rangle_{i\omega_1-i\nu}
   - \langle \langle f_{\sigma}|
  n_{-\sigma}f^{\dagger}_{\sigma}
  \rangle \rangle_{i\omega_n+i\nu}
  \langle \langle f_{-\sigma}|f_{\sigma}f^{\dagger}_{\sigma}
  f^{\dagger}_{-\sigma} \rangle \rangle_{i\omega_1-i\nu} \biggr)
\nonumber\\
&\times&  \langle \langle f_{-\sigma}f_{\sigma}f^{\dagger}_{\sigma}|
  c^{\dagger}_{{\bf k}-\sigma} \rangle \rangle_{i\omega_1}
+ \frac{\langle f_{\sigma} f^{\dagger}_{\sigma} c^{\dagger}_{{\bf k}-\sigma}
f_{-\sigma} \rangle}
{1 - \langle n_{-\sigma} \rangle}
\langle \langle f_{\sigma} n_{-\sigma}|
   f^{\dagger}_{\sigma} \rangle \rangle_{i\omega_n}
   \label{APPR2}
\end{eqnarray}
and the Green function $G_{3\sigma},G_{4\sigma}$ are decoupled according to the theorem
of Wick. Since the approximation does not violate the theorem of Wick for
small $U$, it automatically satisfies the SOPT, i.e., expanding Eq.
(\ref{EQQ1}) for small $U$ up to second order in $U$ leads to the SOPT for
the self-energy. Also the $V^2$- limit is not violated since the Green function
$G_{3\sigma},G_{4\sigma}$ are itselves proportional to $V^2$, leading in Eq.
(\ref{EQQ}) to $V^4$ terms. Therefore, our approximation leads to an
expression for the self-energy of the SIAM, which is exact at least up to
order $U^2$ in a weak coupling expansion and up to order $V^2$ in a strong
coupling expansion.  The structure of the chosen approximation
(\ref{APPR1}) and (\ref{APPR2}) and of the decoupling for the Green function
$G_{3\sigma},G_{4\sigma}$ according to the theorem of Wick has a similar
analytical structure as the SOPT, (which can be calculated numerically very
fast and accurate). Hence the explicit numerical calculations
within this treatment are of the same order of complexity as those of the
self-consistent-SOPT calculations.\\
Notice that in principle it is possible to systematically improve
the above approximation.
Since the self-consistent summation (\ref{EQQ1}), (\ref{EQQ}) is formally
exact, the next step would be the similar construction of an
approximation for the Green functions
$G_{3\sigma},G_{4\sigma}$ (and for Green functions of a similar structure occurring in a
further application of the equations of motion to the Green functions $G_{1\sigma},G_{2\sigma}$)
being exact in order $V^2$ and
simultaneously satisfying the theorem of Wick;
as the Green functions $G_{3\sigma}$ etc. have already a prefactor
$V^2$ in (\ref{EQQ}) this leads to an approximation for ${\cal S}$ and
thus the self-energy $\Sigma_{\sigma}(\omega)$ being exact up to order $V^4$
in the strong-coupling limit and simultaneously in order $U^2$ in the
weak-coupling limit. Furthermore, already from the structure of the
exact equation (\ref{EQQ1}) it is clear that our new approximation can
be considered as a systematic improvement of the
\emph{Hubbard-III}  approximation (\ref{HUBIII}),
which is known to be reasonable concerning
the high-frequency behavior of the dynamical quantities and concerning
the reproduction of the metal-insulator transition in the Hubbard model.\\ 
The improved
approach goes beyond the \emph{Hubbard-III} approximation~\cite{hubb3} including all
self-energy contributions in order $U^2$ and thus reproducing the SOPT.
This is important to fulfill the Fermi liquid properties at least for small $U$,
and in this respect the approach should be as good as the related attempts.\\
On the other hand, the new approach is
also exact up to order $V^2$ and is, therefore,
as good as standard equations of motion decouling procedures 
are, which qualitatively describe important items like Kondo
peak, Kondo temperature scale, etc.\\
When interpreting these standard equations of motion
decouplings as generalized mean-field treatments, because the decoupling consists in a replacement of a
higher order Green function by a product of an expectation value with a
lower order Green function, our new approximation can be considered to
be a kind of dynamical mean-field approximation, because the
approximation (\ref{APPR1}), (\ref{APPR2}) consists in the replacement of
a  higher order Green function by combinations of products of
(time-dependent) lower order Green functions.\\
Finally the approach is
not a completely uncontrolled approximation, as it is
exact up to certain orders ($V^2$, $U^2$) of systematic perturbation
theory. It is, however, as any self-consistent approximate treatment is, uncontrolled
in the way it takes into account infinite order resummations of
arbitrary order in $U$ and $V$ by the self-consistent requirement, which is
unavoidable to reproduce both limits.\\
In summary, an improved interpolating approximation for the SIAM has been
developed, which recovers the exactly solvable limits $V=0$ and $U=0$ and
which is even more at least correct up to order $V^2$ in a strong-coupling
expansion and simultaneously up to order $U^2$ in a weak-coupling expansion.
%
%
%%%%%%%%%%%%%%%%%%%%%%%%%%%%%%%%%%%%%%%%%%%%%%%%%%%%%
\section{Quasiparticle  Many-Body Dynamics of PAM}
%%%%%%%%%%%%%%%%%%%%%%%%%%%%%%%%%%%%%%%%%%%%%%%%%%%%%
%
The main drawback of the Hartree-Fock type solution of PAM (\ref{eq.61}) is
that it ignores the correlations of the ""up" and "down"
electrons. In this section, we  will take into account the latter
correlations in a self-consistent way using the IGF method. We
consider the relevant matrix Green function of the form (cf.  (\ref{eq.224}))
\begin{equation}
\label{eq.305}
 \hat G (\omega) =
\begin{pmatrix}
\langle \langle c_{k\sigma}\vert c^{\dagger}_{k\sigma}\rangle \rangle &
\langle \langle c_{k\sigma}\vert f^{\dagger}_{k\sigma}\rangle \rangle \cr \langle \langle f_{k\sigma}\vert
c^{\dagger}_{k\sigma}\rangle \rangle & \langle \langle f_{k\sigma}\vert
f^{\dagger}_{k\sigma}\rangle \rangle \\
\end{pmatrix}.
\end{equation}
The equation of motion for Green function (\ref{eq.305}) reads
\begin{eqnarray}
\label{eq.306} 
\begin{pmatrix}
(\omega - \epsilon_{k})& - V_{k} \\ 
-V_{k} & (\omega - E_{k}) \\
\end{pmatrix} 
\begin{pmatrix} 
\langle \langle c_{k\sigma} \vert
c^{\dagger}_{k\sigma}\rangle \rangle & \langle \langle c_{k\sigma}\vert
f^{\dagger}_{k\sigma}\rangle \rangle \cr \langle \langle f_{k\sigma} \vert
c^{\dagger}_{k\sigma}\rangle \rangle & \langle \langle f_{k\sigma} \vert
f^{\dagger}_{k\sigma}\rangle \rangle \\
\end{pmatrix}
 = \nonumber
\\  \begin{pmatrix} 1&0 \\ 0&1 \\ + 
UN^{-1} \sum_{pq} 
\end{pmatrix}
\begin{pmatrix}
0&0 \cr \langle \langle A \vert c^{\dagger}_{k\sigma}\rangle \rangle& \langle \langle A \vert
f^{\dagger}_{k\sigma}\rangle \rangle \\
\end{pmatrix},
\end{eqnarray}
where
 $ A = f_{k+p\sigma}f^{\dagger}_{p+q-\sigma}f_{q-\sigma}$.
According to IGF method the definition of the irreducible parts in
the equation of motion (\ref{eq.306})  are given by
\begin{eqnarray}
 ^{(ir)}\langle \langle f_{k+p\sigma}f^{\dagger}_{p+q-\sigma}f_{q-\sigma} \vert
c^{\dagger}_{k\sigma}\rangle \rangle =
\langle \langle f_{k+p\sigma}f^{\dagger}_{p+q-\sigma}f_{q-\sigma} \vert
c^{\dagger}_{k\sigma}\rangle \rangle  -  \delta_{p,0}\langle n_{q-\sigma} \rangle
\langle \langle f_{k\sigma}\vert c^{\dagger}_{k\sigma}\rangle \rangle,   \\
^{(ir)}\langle \langle f_{k+p\sigma}f^{\dagger}_{p+q-\sigma}f_{q-\sigma} \vert
f^{\dagger}_{k\sigma}\rangle \rangle =
\langle \langle f_{k+p\sigma}f^{\dagger}_{p+q-\sigma}f_{q-\sigma} \vert
f^{\dagger}_{k\sigma}\rangle \rangle  -  \delta_{p,0} \langle n_{q-\sigma} \rangle
\langle \langle f_{k\sigma}\vert f^{\dagger}_{k\sigma}\rangle \rangle.  
\end{eqnarray}
After substituting   these definitions into   equation
(\ref{eq.306}), we obtain
\begin{eqnarray}
\label{eq.307} 
\begin{pmatrix}
(\omega - \epsilon_{k})& - V_{k} \\ 
- V_{k} & (\omega - E_{\sigma}(k)) \\
\end{pmatrix}
\begin{pmatrix} \langle \langle c_{k\sigma}
\vert c^{\dagger}_{k\sigma}\rangle \rangle & \langle \langle c_{k\sigma}\vert
f^{\dagger}_{k\sigma}\rangle \rangle \cr \langle \langle f_{k\sigma} \vert
c^{\dagger}_{k\sigma}\rangle \rangle & \langle \langle f_{k\sigma} \vert
f^{\dagger}_{k\sigma}\rangle \rangle \\ 
\end{pmatrix}
= \nonumber
\\ \begin{pmatrix} 1&0 \\ 0&1 \\
\end{pmatrix}
 + UN^{-1} \sum_{pq} 
 \begin{pmatrix}
0&0 \\ ^{(ir)}\langle \langle A \vert c^{\dagger}_{k\sigma}\rangle \rangle& ^{(ir)}\langle \langle A \vert f^{\dagger}_{k\sigma}\rangle \rangle \\
\end{pmatrix}.
\end{eqnarray}
In the the following the notation  will be used for brevity
\begin{equation}
E_{\sigma}(k) = E_{k} - Un_{-\sigma}; \quad n_{-\sigma} = \langle f^{\dagger}_{k-\sigma}f_{k-\sigma}\rangle.
\end{equation}
The definition of the generalized mean field Green function (which, for the weak Coulomb
correlation $U$, coincides with the Hartree-Fock mean field ) is
evident. All inelastic renormalization terms are now related to
the last term in the equation of motion (\ref{eq.307}). All
elastic scattering ( or mean field) renormalization terms are
included into the following  mean-field Green function
\begin{equation}
\label{eq.308} 
\begin{pmatrix}
(\omega - \epsilon_{k})& - V_{k} \\ -
V_{k} & (\omega - E_{\sigma}(k)) \\ 
\end{pmatrix}
\begin{pmatrix} 
\langle \langle c_{k\sigma}\vert c^{\dagger}_{k\sigma}\rangle \rangle^{0} & \langle \langle c_{k\sigma}\vert
f^{\dagger}_{k\sigma}\rangle \rangle^{0} \\ \langle \langle f_{k\sigma} \vert
c^{\dagger}_{k\sigma}\rangle \rangle^{0} & \langle \langle f_{k\sigma} \vert
f^{\dagger}_{k\sigma}\rangle \rangle^{0} \\
\end{pmatrix}
= \nonumber\\  
\begin{pmatrix} 
1&0 \\ 0&1 \\
\end{pmatrix}.
\end{equation}
It is easy to find that ({\it cf.} (\ref{eq.229}) and
(\ref{eq.230}))
\begin{eqnarray}
\label{eq.309} \langle \langle f_{k\sigma} \vert f^{\dagger}_{k\sigma}\rangle \rangle^{0} =
\Bigl ( \omega - E_{\sigma}(k) - \frac { |V_{k}|^{2}}{\omega -
\epsilon_{k}} \Bigr )^{-1}, \\
\label{eq.310}
 \langle \langle c_{k\sigma} \vert c^{\dagger}_{k\sigma}\rangle \rangle^{0} =
 \Bigl
( \omega - \epsilon_{k} - \frac { |V_{k}|^{2}}{\omega -
E_{\sigma}(k)} \Bigr )^{-1}.
\end{eqnarray}
At this point, it is worthwhile to emphasize a significant
difference between  both the models, PAM and SIAM. The
corresponding SIAM equation for generalized mean field Green function
(\ref{eq.228}) reads
\begin{eqnarray}
\label{eq.311} \sum_{p}  
\begin{pmatrix}
(\omega -
\epsilon_{p})\delta_{pk}& - V_{p}\delta_{pk} \\ - V_{p} & {1
\over N} (\omega - E_{0\sigma} - Un_{-\sigma}) \\
\end{pmatrix}
\begin{pmatrix}
\langle \langle c_{k\sigma} \vert c^{\dagger}_{k\sigma}\rangle \rangle^{0} &
\langle \langle c_{k\sigma}\vert f^{\dagger}_{0\sigma}\rangle \rangle^{0} \\ \langle \langle f_{0\sigma}
\vert c^{\dagger}_{k\sigma}\rangle \rangle^{0} & \langle \langle f_{0\sigma} \vert
f^{\dagger}_{0\sigma}\rangle \rangle^{0}\\
\end{pmatrix} 
= \nonumber \\
 \begin{pmatrix} 
 1&0 \\ 0&1 \\
 \end{pmatrix}.
\end{eqnarray}
This   matrix notation for SIAM shows a fundamental distinction
between SIAM and PAM. For SIAM,  we have a different number of
states for a strongly localized level and the conduction electron
subsystem: the conduction band contains $2N$ states, whereas the
localized (s-type) level contains only two. The comparison of
(\ref{eq.311}) and (\ref{eq.308}) shows clearly that this
difficulty does not exist for PAM : the number of states   both
in the localized and itinerant subsystems are the same,   i.e.  $2N$.\\
This important difference between SIAM and PAM   appears also
when we calculate   inelastic scattering or self-energy
corrections. By analogy with the Hubbard model~\cite{kuz78,kuznc94,kuzrnc02}, the equation of
motion (\ref{eq.307}) for PAM can be transformed exactly to the
scattering equation of the form (\ref{e77}). Then, we are able to write
down explicitly the Dyson equation (\ref{e78}) and the exact expression
for the self-energy M  in the matrix form:
\begin{equation}
\label{eq.312} \hat M_{k\sigma}(\omega) = 
\begin{pmatrix}
0&0\\ 0&M_{22}\\
\end{pmatrix}.
\end {equation}
Here the matrix element $ M_{22}$ is of the form
\begin{eqnarray}
\label{eq.313} M_{22} = M_{k\sigma}(\omega) = \\
\frac{U^2}{N^2}{} \sum_{pqrs}{} \bigl (^{(ir)}\langle \langle f_{k+p\sigma}f^{\dagger}_{p+q-\sigma}
f_{q-\sigma} \vert
f^{\dagger}_{r-\sigma}f_{r+s-\sigma}f^{\dagger}_{k+s\sigma}\rangle \rangle^{(ir)}
\bigr )^{(p)}. \nonumber
\end{eqnarray}
To calculate the self-energy operator (\ref{eq.313}) in a
self-consistent way, we proceed by analogy with the Hubbard model. Then we find both the expressions for the self-energy
operator~\cite{kuz78,kuznc94,kuzrnc02} by iteration procedure. 
%in   form (\ref{eq.149}) and (\ref{eq.152}).
%
%
%%%%%%%%%%%%%%%%%%%%%%%%%%%%%%%%%%%%%%%%%%%%%%%%%%%%%%%%%%%
\section{Quasiparticle Many-Body Dynamics of TIAM}
%%%%%%%%%%%%%%%%%%%%%%%%%%%%%%%%%%%%%%%%%%%%%%%%%%%%%%%%%%%
%
%
Let us see now how to apply the results of the preceding
Sections for the case of TIAM Hamiltonian (\ref{eq.62}). 
The initial intention of  Alexander and Anderson~\cite{ander64} was to extend
the theory of localized magnetic states of solute atoms in metals to the case of a pair of neighboring magnetic atoms~\cite{kpb93,gavo81}. 
It was found that the simplified model based on the idea that the important interaction is the diagonal exchange integral 
in the localized state, which is exactly soluble in Hartree-Fock theory for isolated ions, is still soluble, and the 
solutions show both ferromagnetic and antiferromagnetic exchange mechanisms.\\
Contrary to that, our approach go beyond the Hartree-Fock approximation and permits one to describe the quasiparticle
many-body dynamics of TIAM in a self-consistent way.\\
We again consider the relevant matrix Green function of the form (cf.(\ref{eq.224}))
\begin{equation}
\label{eq.314}
 \hat G (\omega) = 
 \begin{pmatrix}
 G_{11}&G_{12}&G_{13} \\
 G_{21}&G_{22}&G_{23} \\
 G_{31}&G_{32}&G_{33}\\
\end{pmatrix} 
  =
\begin{pmatrix} 
\langle \langle c_{k\sigma}\vert c^{\dagger}_{k\sigma}\rangle \rangle &
\langle \langle c_{k\sigma}\vert f^{\dagger}_{1\sigma}\rangle \rangle&  \langle \langle c_{k\sigma}\vert
f^{\dagger}_{2\sigma}\rangle \rangle \cr \langle \langle f_{1\sigma}\vert
c^{\dagger}_{k\sigma}\rangle \rangle &\langle \langle f_{1\sigma}\vert
f^{\dagger}_{1\sigma}\rangle \rangle& \langle \langle f_{1\sigma}\vert
f^{\dagger}_{2\sigma}\rangle \rangle \cr \langle \langle f_{2\sigma}\vert
c^{\dagger}_{k\sigma}\rangle \rangle &\langle \langle f_{2\sigma}\vert
f^{\dagger}_{1\sigma}\rangle \rangle& \langle \langle f_{2\sigma}\vert
f^{\dagger}_{2\sigma}\rangle \rangle \\
\end{pmatrix}.
\end{equation}
The equation of motion for Green function (\ref{eq.314}) reads
\begin{eqnarray}
\label{eq.315} \sum_{p} 
\begin{pmatrix}
(\omega - \epsilon_{p})\delta_{pk}& - V_{1p}\delta_{pk}& -
V_{1p}\delta_{pk} \\ - V_{1p} & {1 \over N} (\omega - E_{0\sigma}
)& -V_{12} \\ -V_{2p} & - V_{21} & {1 \over N} (\omega -
E_{0\sigma}) \\
\end{pmatrix}
 \begin{pmatrix}
 G_{11}&G_{12}&G_{13} \\
 G_{21}&G_{22}&G_{23} \\
 G_{31}&G_{32}&G_{33}\\
 \end{pmatrix}
= \nonumber \\
 \begin{pmatrix} 
 1&0&0 \cr 0&1&0 \cr
 0&0&1 \\
 \end{pmatrix}
 +  U
\begin{pmatrix}
0&0&0 \cr \langle \langle A_{1} \vert c^{\dagger}_{k\sigma}\rangle \rangle & \langle \langle 
A_{1}\vert f^{\dagger}_{1\sigma}\rangle \rangle & \langle \langle A_{1} \vert
f^{\dagger}_{2\sigma}\rangle \rangle \cr \langle \langle A_{2} \vert c^{\dagger}_{k\sigma}\rangle \rangle
& \langle \langle A_{2}\vert f^{\dagger}_{1\sigma}\rangle \rangle & \langle \langle A_{2} \vert
f^{\dagger}_{2\sigma}\rangle \rangle \\
\end{pmatrix}.
\end{eqnarray}
The notation  are as follows 
\begin{equation}
 A_{1} = f_{1\sigma}f^{\dagger}_{1-\sigma}f_{1-\sigma}; \quad  A_{2} =
f_{2\sigma}f^{\dagger}_{2-\sigma}f_{2-\sigma}. 
\end{equation}
In a compact notation, the equation (\ref{eq.315}) has the form  
\begin{equation}
\label{eq.316} \sum_{p}F(p,k)G_{pk}(\omega) = \hat I + U
D_{p}(\omega).
\end{equation}
We thus have the equation of motion (\ref{eq.316}) which is a
complete analogue of the corresponding equations for the SIAM and
PAM. After introducing the irreducible parts by analogy with the
equation (\ref{eq.225})
\begin{eqnarray}
 ^{(ir)}\langle \langle f_{1\sigma}f^{\dagger}_{1-\sigma}f_{1-\sigma}
\vert B\rangle \rangle_ {\omega} =
\langle \langle f_{1\sigma}f^{\dagger}_{1-\sigma}f_{1-\sigma}\vert B\rangle \rangle_{\omega}  
  - \langle n_{1-\sigma} \rangle \langle \langle f_{1\sigma} \vert B\rangle \rangle_{\omega},\\
^{(ir)}\langle \langle f_{2\sigma}f^{\dagger}_{2-\sigma}f_{2-\sigma}
\vert B\rangle \rangle_ {\omega} =
\langle \langle f_{2\sigma}f^{\dagger}_{2-\sigma}f_{2-\sigma}\vert B\rangle \rangle_{\omega}   
  - \langle n_{2-\sigma} \rangle \langle \langle f_{2\sigma} \vert B\rangle \rangle_{\omega},\nonumber 
\end{eqnarray}
and performing the second-time   differentiation of the
higher-order Green function, and introducing the relevant irreducible parts,
the equation of motion (\ref{eq.316}) is rewritten  in the form of
Dyson equation (\ref{e78}). The definition of the generalized mean field Green function  is as follows
\begin{eqnarray}
\label{eq.317} \sum_{p} 
\begin{pmatrix}
(\omega - \epsilon_{p})\delta_{pk}& - V_{1p}\delta_{pk}& -
V_{1p}\delta_{pk} \\ - V_{1p} & {1 \over N} (\omega -
E_{0\sigma} - Un_{-\sigma} )& -V_{12} \cr -V_{2p} & - V_{21} & {1
\over N} (\omega - E_{0\sigma} - Un_{-\sigma}) \\
\end{pmatrix} \times
 \nonumber \\
\begin{pmatrix}
G^{0}_{11}&G^{0}_{12}&G^{0}_{13} \\
 G_{21}&G^{0}_{22}&G^{0}_{23} \\
 G^{0}_{31}&G^{0}_{32}&G^{0}_{33}\\
 \end{pmatrix}
  =
\begin{pmatrix}
 1&0&0 \cr 0&1&0 \\
 0&0&1 \\
\end{pmatrix}.
\end{eqnarray}
The matrix Green function (\ref{eq.317}) describes the mean-field solution of
the TIAM Hamiltonian. The explicit solutions for diagonal
elements of $G^{0}$ are 
\begin{eqnarray}
\label{eq.318}   \langle \langle c_{k\sigma} \vert
c^{\dagger}_{k\sigma}\rangle \rangle^{0}_{\omega} =   \Bigl ( \omega
-\epsilon_{k} - \frac { |V_{1k}|^{2}}{\omega - (E_{0\sigma} -
Un_{-\sigma})} - \Delta_{11}(k,\omega)  \Bigr )^{-1}, \\
 \label{eq.319} \langle \langle f_{1\sigma} \vert f^{\dagger}_{1\sigma}\rangle \rangle^{0}_
{\omega} = \Bigl ( \omega - ( E_{0\sigma} - Un_{-\sigma}) -
S (\omega)) - \Delta_{22}(k,\omega) \Bigr )^{-1}, \\
\label{eq.320} \langle \langle f_{2\sigma} \vert f^{\dagger}_{2\sigma}\rangle \rangle^{0}_
{\omega} = \Bigl ( \omega - ( E_{0\sigma} - Un_{-\sigma}) - S
(\omega)) - \Delta_{33}(k,\omega) \Bigr )^{-1}.
\end{eqnarray}
Here we   introduced the notation
\begin{align}
\label{eq.321} \nonumber \Delta_{11}(k,\omega) = \Bigl( V_{2k} +
\frac { V_{1k}V_{12}}{\omega - (E_{0\sigma} - Un_{-\sigma})}
\Bigr) \Bigl( V_{2k} + \frac { V_{1k}V_{21}}{\omega - (E_{0\sigma}
- Un_{-\sigma})} \Bigr) \times
\\ \Bigl [ \omega - (E_{0\sigma} - Un_{-\sigma}) - \frac {
V_{21}V_{12}}{\omega - (E_{0\sigma} - Un_{-\sigma})} \Bigr ]^{-1},  \\
 \Delta_{22}(k,\omega) = ( \lambda_{21} (\omega) +
V_{12} )( \lambda_{21} (\omega) + V_{21} ) \Bigl [\omega -
(E_{0\sigma}-  Un_{-\sigma})
- \frac {\sum_{p} |V_{2p}|^{2}}{\omega - \epsilon_{p}} \Bigr ]^{-1},  \\
  \Delta_{33}(k,\omega) = ( \lambda_{12} (\omega) +
V_{21} )( \lambda_{12} (\omega) + V_{12} ) \Bigl [\omega - (E_{0\sigma}
-  Un_{-\sigma}) - \frac
{\sum_{p} |V_{1p}|^{2}}{\omega - \epsilon_{p}} \Bigr ]^{-1}, \\
\lambda_{12} = \lambda_{21} = \sum_{p} \frac {V_{1p}V_{2p}}{\omega
- \epsilon_{p}}.
\end{align}
The formal solution of the Dyson equation for TIAM contains the
self-energy matrix
\begin{equation}
\label{eq.322} \hat M = 
\begin{pmatrix}
0&0&0\cr
 0&M_{22}&M_{23} \\
 0&M_{32}&M{33}\\
\end{pmatrix} ,
\end{equation}
where
\begin{eqnarray}
\label{eq.323} M_{22} =  U^{2}
(^{(ir)}\langle \langle f_{1\sigma}n_{1-\sigma}\vert
f^{\dagger}_{1\sigma}n_{1-\sigma}\rangle \rangle^{(ir)})^{p}, \\  
M_{32} = U^{2} (^{(ir)}\langle \langle f_{2\sigma}n_{2-\sigma}\vert
f^{\dagger}_{1\sigma}n_{1-\sigma}\rangle \rangle^{(ir)})^{p}, \\  
M_{23} = U^{2} (^{(ir)}\langle \langle f_{1\sigma}n_{1-\sigma}\vert
f^{\dagger}_{2\sigma}n_{2-\sigma}\rangle \rangle^{(ir)})^{p}, \\  
M_{33} = U^{2} (^{(ir)}\langle \langle f_{2\sigma}n_{2-\sigma}\vert
f^{\dagger}_{2\sigma}n_{2-\sigma}\rangle \rangle^{(ir)})^{p}.
\end{eqnarray}
To calculate the matrix elements (\ref{eq.323}), the same
procedure can be used as it was done previously for the SIAM
(\ref{eq.239}). As a result, we find the following explicit
expressions for the self-energy matrix elements (cf.(\ref{eq.241})
\begin{eqnarray}
\label{eq.324} M^{\uparrow}_{22}(\omega) = U^2
\int_{-\infty}^{+\infty} d\omega_{1}d{\omega}_{2}\frac{1 +
N(\omega_{1}) - n(\omega_{2})}
{\omega - \omega_{1} - \omega_{2}} \times \nonumber\\
\Bigl ( -{1 \over \pi}Im \langle \langle S^{-}_{1} \vert S^{+}_{1}\rangle \rangle_{\omega_{1}} \Bigr) 
\Bigl ( -{1 \over \pi}Im \langle \langle f_{1\downarrow}\vert
f^{\dagger}_{1\downarrow}\rangle \rangle_{\omega_{2}} \Bigr),  \\
\label{eq.325} M^{\downarrow}_{22}(\omega) = U^2
\int_{-\infty}^{+\infty} d\omega_{1}d{\omega}_{2}\frac{1 +
N(\omega_{1}) - n(\omega_{2})}
{\omega - \omega_{1} - \omega_{2}} \times \nonumber\\
\Bigl ( -{1 \over \pi}Im \langle \langle S^{+}_{1} \vert S^{-}_{1}\rangle \rangle_{\omega_{1}} \Bigr ) 
\Bigl ( -{1 \over \pi}Im \langle \langle f_{1\uparrow}\vert
f^{\dagger}_{1\uparrow}\rangle \rangle_{\omega_{2}} \Bigr ),  \\
\label{eq.326} M^{\uparrow}_{23}(\omega) = U^2
\int_{-\infty}^{+\infty} d\omega_{1}d{\omega}_{2}\frac{1 +
N(\omega_{1}) - n(\omega_{2})}
{\omega - \omega_{1} - \omega_{2}} \times \nonumber\\
\Bigl ( -{1 \over \pi}Im \langle \langle S^{-}_{1} \vert S^{+}_{2}\rangle \rangle_{\omega_{1}} \Bigr ) 
\Bigl ( -{1 \over \pi}Im\langle \langle f_{1\downarrow}\vert
f^{\dagger}_{2\downarrow}\rangle \rangle_{\omega_{2}} \Bigr ),  \\
\label{eq.327} M^{\downarrow}_{23}(\omega) = U^2
\int_{-\infty}^{+\infty} d\omega_{1}d{\omega}_{2}\frac{1 +
N(\omega_{1}) - n(\omega_{2})}
{\omega - \omega_{1} - \omega_{2}} \times \nonumber\\
\Bigl ( -{1 \over \pi}Im \langle \langle S^{+}_{2} \vert S^{-}_{1}\rangle \rangle_{\omega_{1}} \Bigr ) 
\Bigl ( -{1 \over \pi}Im\langle \langle f_{1\uparrow}\vert
f^{\dagger}_{2\uparrow}\rangle \rangle_{\omega_{2}} \Bigr ).
\end{eqnarray}
Here the following notation  were used:
$$S^{+}_{i} = f^{\dagger}_{i\uparrow}f_{i\downarrow}; \quad
S^{-}_{i} = f^{\dagger}_{i\downarrow}f_{i\uparrow}; \quad i=1,2.$$  
For $ M_{33}$ we obtain the same expressions as for
$M_{22}$  with the substitution of index 1 by 2.  For
$M^{\uparrow \downarrow}_{32}$ we must do the same. It is
possible to say that the diagonal elements $M_{22}$ and $M_{33}$
describe single-site inelastic scattering processes; off-diagonal
elements $M_{23}$ and $M_{32}$ describe   intersite inelastic
scattering processes. They are responsible for the specific
features of the dynamic behavior of TIAM ( as well as the off-diagonal matrix elements of the Green function
$G^{0}$) and, more generally, the cluster impurity Anderson model
(CIAM). The nonlocal contributions to the total spin
susceptibility of two well formed impurity magnetic moments at a
distance $R$ can be estimated as
\begin{equation}
\chi_{pair} \sim \langle \langle S^{-}_{1} \vert S^{+}_{2}\rangle \rangle \sim 2\chi -12\pi
E_{F}(\frac {\chi}{ g \mu_{B}})^{2} \frac {\cos
(2k_{F}R)}{(k_{F}R)^{3}}.
\end{equation}
In the region of interplay of the RKKY and Kondo behavior, the
key point is then to connect the partial Kondo screening effects
with the low temperature behavior of the total spin
susceptibility. As it is known, it is quite difficult to describe
such a threshold behavior analytically. However, progress is
expected due to a better understanding of the quasiparticle
many-body dynamics both from analytical and numerical investigations.
%
%
%
%%%%%%%%%%%%%%%%%%%%%% 
\section{Conclusions}
%%%%%%%%%%%%%%%%%%%%%% 
%

In summary, we presented in this paper in terse form  a general technique how  a
dynamical solution for SIAM and TIAM at finite temperatures and for the
broad interval of the values of the model parameters can be constructed
in the spirit of irreducible Green  functions approach. We used an exact
result to connect the single-particle Green  function with the higher-order Green  function to
obtain an complex expansion in terms of $U$ and $V$ for the propagator.
This approach provides a  plausible yet sound understanding of how structure 
of the relevant dynamical solution may be found.
Hence this  approach offer a both powerful and workable technique for a systematic construction  of
the approximative dynamical solutions of SIAM, PAM and other models
of the strongly correlated electron systems. \\
In short, the theory of the many-body quasiparticle dynamics
of the Anderson-  and Hubbard-type models at finite temperatures
have been reviewed. We stressed an importance  of the new \textbf{exact identity} relating the
one-particle and many-particle Green functions for the single-impurity Anderson
model: $G = g_{0} + g_{0}Pg_{0}$.\\
The application of the IGF method to the investigation of nonlocal
correlations and quasiparticle interactions in Anderson
models~\cite{kpb93} has a particular interest for  studying of the
inter-site correlation effects in the concentrated Kondo system and other problems of solid state physics~\cite{gavo81,davyd12}. A
comparative study of real many-body dynamics of single-impurity,
two-impurity, and periodic Anderson model, especially for strong
but finite Coulomb correlation, when perturbation expansion in
$U$ does not work, is  of importance  for the characterization  of the true
quasiparticle excitations and the role of magnetic correlations.
It was shown that the physics of two-impurity Anderson model can
be understood in terms of competition between   itinerant motion
of carriers and magnetic correlations of the RKKY nature. This
issue is still very controversial and the additional efforts must
be applied in this field.\\ 
The many-body quasiparticle dynamics of the
single-impurity Anderson Model was investigated by means of the
equations of motion for the higher-order Green functions. It
was shown that an interpolating approximation, which simultaneously
reproduces the weak-coupling limit up to second order in the
interaction strength $U$ and the strong coupling limit up to
second order in the hybridization $V$ (and thus also fulfills
the atomic limit) may be formulated self-consistently. Hence, a new advanced 
 many-body dynamical solution for SIAM has
been developed, which recovers the exactly solvable limits $V=0$ and
$U=0$  and which is even more at least correct up to order
$V^2$ in a strong-coupling expansion and simultaneously up to order $U^2$ in a
weak-coupling expansion.\\
Further applications and development of the technique of the equations of motion for the Green functions
were described in Refs.~\cite{gors11,petr93,wang99,schaf99,alas04,kogo06,gors13,miz13,miz14,gors14,ashok15}
These applications illustrate some of   subtle details of this  
approach and exhibit the  physical significance and operational ability of the Green function technique in 
a representative  form.\\ 
This line of consideration is very
promising for developing the complete and self-contained theory
of strongly interacting many-body systems on 
a lattice~\cite{kuz09,kuzapp76,kuzam91,kpb93,kuz94,ckw96,kuzem96,kuapp97,kuz78,igf89,kuznc94,kuzmpl97,kuzrnc02,kuzcmp10,kuzsf99,kuz04,dms05}.
Our main results reveal the fundamental importance of the adequate
definition of \emph{generalized mean fields} at finite temperatures, that
results in a   deeper insight into the nature of quasiparticle
states of the correlated lattice fermions and spins. We believe
that our approach offers a new way for   systematic constructions
of the approximate dynamic  solutions of the Hubbard, SIAM, TIAM,
PAM, spin-fermion, and other models of the strongly correlated electron systems on a lattice. 
%

%
%
%\section*{References}


\begin{thebibliography}{99}


%% \bibitem must have the following form:
%%   \bibitem{key}...
%%

% \bibitem{}
%
%
%
% 
% 
\bibitem{ander61}
P. W. Anderson,  Localized magnetic states in metals.
\emph{Phys. Rev}. \textbf{124}, 41 (1961). 
%
%
%
\bibitem{and79}
P. W. Anderson, Local moments and localized states. \emph{Rev. Mod. Phys.} \textbf{50}, 191 (1978).   
%
%
%
\bibitem{ander64}
S. Alexander  and P. W. Anderson, Interaction between localized states in metals.
\emph{Phys. Rev.} \textbf{A133}, 1594 (1964). 
%
%
%
%
\bibitem{tyab}  S. V. Tyablikov, \emph{Methods in the Quantum Theory of Magnetism}
(Plenum Press, New York, 1967).
%
%
%
%
\bibitem{kuz09}
A. L. Kuzemsky, Statistical mechanics and the physics of  many-particle model systems.  
{\em Physics of  Particles and  Nuclei},  {\bf 40}, 949 (2009); [arXiv: [cond-mat.str-el] 1101.3423].
%
% 
%
%
\bibitem{kuzapp76}
A. L. Kuzemsky,
To the correlation theory of d-electrons in transition metals.  \emph{Acta Phys. Polon}. A \textbf{49}, 169-180 (1976).
%

%
\bibitem{kuzam91}
A. L. Kuzemsky,
Interpolation solution of the single-impurity Anderson model.
\emph{Phys. Lett.}  A \textbf{153},   466   (1993).   
%
%
%
%
\bibitem{kpb93}
A. L. Kuzemsky, J. C. Parlebas and H. Beck,
Non-local correlations and quasiparticle interactions in the Anderson model.
\emph{Physica}  A \textbf{198},   606   (1993). 
%
%
%
%
\bibitem{kuz94}  A. L. Kuzemsky,  Correlation effects in high-temperature superconductors and heavy fermion compounds, in:  
\emph{Superconductivity and Strongly Correlated Electron Systems},    C. Noce, A. Romano and  G. Scarpetta  (eds.) 
(World Scientific, Singapore, 1994), p.346-376.   
%
%
%
%
\bibitem{ckw96}
G. Czycholl,  A. L. Kuzemsky,  S. Wermbter,   New interpolative treatment of the single-impurity Anderson model.
\emph{Europhys. Lett}.  \textbf{34}   133-138  (1996).
%
%
%
%
\bibitem{kuzem96}
A. L. Kuzemsky,  Quasiparticle many-body dynamics of the Anderson model.
\emph{Int. J. Mod. Phys.} B \textbf{10}, 1895-1912  (1996).  
%
%
%
\bibitem{kuapp97}
A. L. Kuzemsky,
Spectral properties and new interpolative dynamical solution of the Anderson model.
\emph{Acta Phys. Polon}. A \textbf{92}, 355-358  (1997). 
%
%
%
\bibitem{kuzem05}
A. L. Kuzemsky, Physics of complex magnetic materials: quasiparticle many-body dynamics.
[arXiv:cond-mat/0512183].
%
% 
%
%
%
\bibitem{bech15}
F. Bechstedt,
\emph{Many-Body Approach to Electronic Excitations: Concepts and Applications}   
(Springer,  Berlin, 2015).
%
%
%
\bibitem{rubi00}
P. M. Echenique, J. M. Pitarke, E. V. Chulkov  and  A. Rubio,
Theory of inelastic lifetimes of low-energy electrons in metals.
\emph{Chemical Physics}  \textbf{251}, 1-35 (2000).
%
%
%
%
\bibitem{hoff88}
R. Hoffmann,  
\emph{Solid and Surface: A Chemist's View of Bonding in Extended Structures}    
(Wiley, New York, 1988).   
%
%
%
%
\bibitem{kiko94}
K. A. Kikoin and   V. N. Fleurov,  
\emph{Transition Metal Impurities in Semiconductors: Electronic Structure and Physical Properties}    
(World Scientific, Singapore, 1994).   
%
%
%
\bibitem{dona97}
D. A. McQuarrie,  
\emph{Physical Chemistry: A Molecular Approach}    
(University Science Books, 1997).   
%    
%
%
%
\bibitem{adam97}
A. W. Adamson and A. P. Gast,  
\emph{Physical Chemistry of Surfaces}    
(Wiley, New York, 1997).   
%
%  
%
\bibitem{atki09}
P. Atkins and J. de Paula,  
\emph{Physical Chemistry},  9th edn.  
(W. H. Freeman, New York, 2009).   
%
%
%
\bibitem{kahir03}
E. Kaxiras,
\emph{Atomic and Electronic Structure of Solids}   
(Cambridge University Press, Cambridge,  2003).
%
%
%
%
\bibitem{mar04}
R. M. Martin, \emph{Electronic Structure: Basic Theory and Practical Methods} (Cambridge University Press, Cambridge, 2004).
%
%
%
%
\bibitem{huang08}
P. Huang and E. A. Carter,
Advances in correlated electronic structure methods for solids, surfaces, and nanostructures.
\emph{Annu. Rev. Phys. Chem.}   \textbf{59}, 261-90 (2008).
%
%
%
%
%
\bibitem{kudr84}
V. Drchal  and J. Kudrnovsky,
Electron correlations in alloys of simple and transition metals.
\emph{J. Phys. Chem. Sol.}   \textbf{45}, 267-274 (1984).
%
%
%
%
\bibitem{canad12}
E. Canadell, M.-L. Doublet and C. Iung,
\emph{Orbital Approach to the Electronic Structure of Solids}   
(Oxford University Press, Oxford,  2012).
%
% 
%
%
\bibitem{hubb1}
J. Hubbard, Electron correlations in narrow energy bands.   \emph{Proc. Roy. Soc}. A \textbf{276}, 238 (1963).
%
%
%
%
\bibitem{hubb2}
J. Hubbard, Electron correlations in narrow energy bands. II. The degenerate band case.  
\emph{Proc. Roy. Soc}. A \textbf{277}, 237 (1964).
%
%
%
\bibitem{hubb3}
J. Hubbard, Electron correlations in narrow energy bands. III. An improved solution.   \emph{Proc. Roy. Soc}. A \textbf{281}, 41 (1964).
%
%
%
\bibitem{hubb4}
J. Hubbard, Electron correlations in narrow energy bands. IV. The atomic representation.   \emph{Proc. Roy. Soc}. 
A \textbf{285}, 542 (1965).
%
%
%
\bibitem{hubb5}
J. Hubbard, Electron correlations in narrow energy bands. V. A perturbation expansion about the atomic limit.   
\emph{Proc. Roy. Soc}. A \textbf{296}, 82 (1966).
%
%
%
\bibitem{hubb6}
J. Hubbard, Electron correlations in narrow energy bands. VI. The connection with many-body perturbation theory.   
\emph{Proc. Roy. Soc}. A \textbf{296}, 100 (1966).
%
%
% 
%
\bibitem{kuz78}
A. L. Kuzemsky,  A self-consistent theory of the electron correlation in the Hubbard model.   
\emph{Teor. Mat. Fiz.}   \textbf{36},   208-223 (1978); [\emph{Theor. Math. Phys}. \textbf{36},   692 (1979)].    
%
%
%
%
\bibitem{mizia07}
J. Mizia and G. Gorski,  \emph{Models of Itinerant Ordering in Crystals} 
(Elsevier, Amsterdam, 2007).
%
%
%
\bibitem{gors11}  G. Gorski and J. Mizia, 
Hubbard III approach with hopping interaction and intersite kinetic correlations.
\emph{Phys. Rev.} B \textbf{83}, 064410 (2011). 
%
%
%
\bibitem{matt65}
J. R. Schrieffer and D. C. Mattis, Localized magnetic moments in dilute metallic alloys: correlation effects.
\emph{Phys. Rev}. \textbf{140}, A1412-A1419 (1965). 
% 
%
%
\bibitem{fis71}
K. H. Fischer, Theory of dilute magnetic alloys. \emph{phys. stat. sol}. b \textbf{46}, 11 (1971).    
%
%
%
%
%
\bibitem{faul72}
D. H. Faulkner and J. W. Schweitzer,
Localized magnetic moments in transition metals and alloys.
\emph{J. Phys. Chem. Sol.}   \textbf{33}, 1685-1696 (1972).
% 
%
%
\bibitem{cole77}
B. R. Coles, 
Transitions from local moment to itinerant magnetism as a function of composition in alloys.
\emph{Physica}  B+C \textbf{91}, 167-169 (1977).
%
%
%
%
%
\bibitem{fis78}
K. H. Fischer,  Dilute magnetic alloys with transition metals as host. \emph{Phys. Rep}. \textbf{47}, 225 (1978).    
%
%
%
%
\bibitem{gavo81} 
G. Gavoille and G. Morel, Local moment stability in itinerant electron magnetism.
\emph{J. Magn. Magn. Mat.}   \textbf{23}, 231-236 (1981).
%

%
%
% 
\bibitem{kuz00}
A. L. Kuzemsky,
Fundamental principles of the physics of magnetism and the problem of itinerant and localized electronic states.
Communication JINR E17-2000-32, Dubna  (2000).   
%
%
%
\bibitem{qprot02}
A. L. Kuzemsky,
 Quantum protectorate and microscopic models of magnetism.
\emph{Int. J. Mod. Phys}.   B \textbf{16}, 803-823 (2002);  
[arXiv: cond-mat/0208222].  
% 
%
%
%
\bibitem{jgood97}
 J. B. Goodenough,
Localized-itinerant electronic transitions in oxides and sulfides.
\emph{Journal of Alloys and Compounds} \textbf{262-263}, 1-9  (1997).
%
%
%
\bibitem{zhou98}
J. B. Goodenough  and J.-S. Zhou
Localized to itinerant electronic transitions in transition-metal oxides with the perovskite structure.
\emph{Chem. Mater.}  \textbf{10}, 2980  (1998).   
%
%
%
%
\bibitem{jgood00}
J. B. Goodenough  and J.-S. Zhou, Localized-itinerant and Mott-Hubbard transitions in several perovskites.
\emph{Journal of Superconductivity} \textbf{13}, 989-993  (2000)
%
%
%
%
\bibitem{good01}
J. B. Goodenough and S.L. Cooper,
\emph{Localized to Itinerant Electronic Transition in Perovskite Oxides},  (Structure and Bonding) vol. 98
(Springer,  Berlin, 2001).
%
%
%
\bibitem{good14}
J. B. Goodenough,
Perspective on engineering transition-metal oxides.
\emph{Chem. Mater.}  \textbf{26}, 820  (2014).
%
%
%
%
\bibitem{capo09}
L. Medici, H. Syed and   M. Capone,  
Genesis of coexisting itinerant and localized electrons in iron pnictides.
\emph{Journal of Superconductivity and Novel Magnetism} \textbf{22}, 535  (2009).
%
%
%
%
\bibitem{veron10}
C. Carbone, M. Veronese, P. Moras, S. Gardonio, C. Grazioli, P. H. Zhou, O. Rader, A. Varykhalov, C. Krull, 
T. Balashov, A. Mugarza, P. Gambardella, S. Lebegue, O. Eriksson, M. I. Katsnelson  and A. I. Lichtenstein.
Correlated electrons step by step: itinerant-to-localized transition of Fe impurities in free-electron metal hosts.
\emph{Phys. Rev. Lett.}  \textbf{104}, 117601 (2010). 
%
%
%
\bibitem{wino14}
E. A. Winograd and L. Medici,  
Hybridizing localized and itinerant electrons: A recipe for pseudogaps.
\emph{Phys.Rev.} B \textbf{89}, 085127 (2014).
%
%

% 
%
\bibitem{gerd86}
G. Czycholl, Approximate treatments of intermediate valence and heavy fermion model systems.
\emph{Phys. Rep}. \textbf{143}, 277 (1986).    
%
%
%
%
\bibitem{rama90}
G. Gangadhar Reddy  and  A. Ramakanth,
On the semiconductor-metal transition in a model for mixed valence systems.
\emph{J. Phys. Chem. Sol.}   \textbf{51}, 515-522 (1990).
%
%
%
%
\bibitem{zhu07}
R. C. Albers and  Jian-Xin Zhu,
Solid-state physics: Vacillating valence.
\emph{Nature}  \textbf{446}, 504  (2007).
%
%
%
\bibitem{steg88}
F. Steglich,
Heavy fermions.
\emph{J. Phys. Chem. Sol.}   \textbf{50}, 225-232 (1989).
% 
%    
%
%
\bibitem{hew93}
A. C. Hewson,  \emph{Kondo Problem to Heavy Fermions}. (Cambridge University Press, Cambridge, 1993).
%
%
%
%
\bibitem{gehr02}
G. A. Gehring,
The Anderson model: why does it continue to be so fascinating?
\emph{J. Phys.: Condens. Matter}  \textbf{14}, V5-V8  (2002).
%
%
%
%
\bibitem{uchi15}
Shin-ichi Uchida,
\emph{High Temperature Superconductivity: The Road to Higher Critical Temperature}   
(Springer,  Berlin, 2015).
%
%
%
\bibitem{rama94}
T. Venkatappa Rao, G. Gangadhar Reddy  and  A. Ramakanth,
Finite temperature magnetic properties of the highly correlated Anderson lattice.
\emph{J. Phys. Chem. Sol.}   \textbf{55}, 175-183 (1994).
% 
%
%
%
\bibitem{ogur02}
A. Oguri,
Low-energy properties of an out of equilibrium Anderson model.
\emph{J. Phys. Chem. Sol.}   \textbf{63}, 1591-1594 (2002).
% 
%
%
\bibitem{mosca96}
V. A. Moskalenko, D. Digor, L. Dogotaru and I. Porcescu,
New approach to periodic Anderson model.
\emph{J. of Low Temp. Phys.}   \textbf{105}, 633 (1996). 
%
%
%
%
%
\bibitem{yang96}
J. L. Whitten and H. Yang,
Theory of chemisorption and reactions on metal surfaces.
\emph{Surface Science Reports} \textbf{24}, 55-124 (1996).
%
%
%
\bibitem{bell76}
B. Bell and A. Madhukar,
Theory of chemisorption on metallic surfaces: Role of intra-adsorbate Coulomb correlation and surface structure.
\emph{Phys.Rev.} B \textbf{14}, 4281 (1976).
%
%
%
\bibitem{bren78}
W. Brenig and K. Schonhammer,
Comment on "Theory of chemisorption of metallic surfaces: Role of intra-adsorbate Coulomb correlation and surface structure".
\emph{Phys.Rev.} B \textbf{17}, 4107 (1978).
%
%
%
\bibitem{gfthe06}
S. G. Davison and K.W. Sulston,
\emph{Green-Function Theory of Chemisorption}   
(Springer,  Berlin, 2006).
%
%
%
%
\bibitem{davyd12}  
S. Yu. Davydov,
The Alexander-Anderson problem for two atoms adsorbed on graphene.
\emph{Physics of the Solid State}   \textbf{54}, 1728 (2012).
%
%
%
%
\bibitem{brod99}
N. B. Zhitenev, M. Brodsky, R. C. Ashoori, L. N. Pfeiffer  and K. W. West,    
Localization-delocalization transition in quantum dots.
\emph{Science}   \textbf{285}, 715 (1999).
%
%
%
%
%
\bibitem{kouw01} 
L. P. Kouwenhoven, D. G. Austing  and S. Tarucha,    Few-electron quantum dots.
\emph{Rep. Prog. Phys.}   \textbf{64}, 701 (2001).
%
%
%
\bibitem{mann02} 
S. M. Reimann  and M. Manninen,    Electronic structure of quantum dots.
\emph{Rev. Mod  Phys.}   \textbf{74}, 1283 (2002).
%
%
%
%
\bibitem{pete04} 
P. Michler (ed.), \emph{Single Quantum Dots: Fundamentals, Applications and New Concepts} 
(Springer, Berlin, 2004)
%
%
%
%
\bibitem{ashok01}
A. Chatterjee and S. Mukhopadhyay, Polaronic effects in quantum dots. 
\emph{Acta Phys. Pol.} B \textbf{32}, 473 (2001).
% 
%
%
%
\bibitem{nazar02}
M. Eto and  Yu. V. Nazarov,
Scaling analysis of the Kondo effect in quantum dots with an even number of electrons.
\emph{J. Phys. Chem. Sol.}   \textbf{63}, 1527-1530 (2002).
% 

%
%
\bibitem{kuo03}
Yia-Chung Chang and David M.-T. Kuo,
Effects of electron correlation on the photocurrent in quantum dot infrared photodetectors.
\emph{Appl. Phys. Lett.}   \textbf{83}, 156 (2003).
%
%
%
%
\bibitem{ashok07}
B. Boyacioglu  M. Saglam  and    A. Chatterjee,
Two-electron bound states in a semiconductor quantum dots with Gaussian   confinement.
\emph{J. Phys.: Condens. Matter}  \textbf{19}, 456217  (2007).
%
%
%
\bibitem{stef09}
P. Stefanski,
Level occupancy anomalies in a double QD system.
\emph{Acta Phys. Pol.} A \textbf{115}, 92 (2009).
%
%
%
\bibitem{citro09} 
F. Romeo   and R. Citro,
Adiabatic quantum pumping and rectification effects in interacting quantum dots. 
arXiv:0903.2362v2 [cond-mat.mes-hall] (2009), 
% 
%
%
%
\bibitem{ashok12}
B. Boyacioglu   and A. Chatterjee,
Magnetic properties of semiconductor quantum dots with gaussian confinement.
\emph{Int. J. Mod. Phys.} B \textbf{26}, 1250018  (2012).
%
%
%
\bibitem{kuca14}
G. Gorski, J. Mizia and K. Kucab, 
Alternative equation of motion approach applied to transport through a quantum dot.
arXiv:1412.7047 [cond-mat.mes-hall] (2014).  
%
%
%
%
\bibitem{igf89}
A. L. Kuzemsky,    Irreducible Green's function method in condensed matter physics,
\emph{Doklady AN SSSR}     \textbf{309},  323-326 (1989); [Sov.Phys.Dokl. \textbf{34},  974 (1989)].   
%  
%
%
\bibitem{kuznc94}
A. L. Kuzemsky, Generalized mean fields and quasiparticle interactions in the Hubbard model.  
 \emph{Nuovo Cimento}  B \textbf{109},   829   (1994).  
%
%
%
%
\bibitem{kuzmpl97}
A. L. Kuzemsky,
Quasiparticle many-body dynamics of the highly correlated electronic systems.
\emph{Molecular Physics Reports}  \textbf{17}, 221-246 (1997); [arXiv:cond-mat/9704028].
%
%
%
%
\bibitem{kuzrnc02}
A. L. Kuzemsky, Irreducible Green functions method and many-particle interacting 
 systems on a lattice.   \emph{La Rivista del Nuovo Cimento}    \textbf{25,} 1  (2002); [arXiv:cond-mat/0208219].  
%
%
%
%
\bibitem{kuzcmp10}
A. L. Kuzemsky, 
Quasiaverages, symmetry breaking and irreducible Green
functions method. \emph{Condensed Matter Physics}  \textbf{13}, N 4, p.43001: 1-20  (2010).
%
%
%
\bibitem{kuzsf99}
A. L. Kuzemsky,  Spectral properties of the generalized spin-fermion models.
\emph{Int. J. Mod. Phys.} B \textbf{13}, 2573  (1999).
% 
%
%
\bibitem{kuz04}
A. L. Kuzemsky, 
 Bound and scattering states of itinerant charge carriers in complex magnetic materials.
\emph{Int. J. Mod. Phys.} B \textbf{18}, 3227  (2004). 
%
%
%
%
\bibitem{dms05}
A. L. Kuzemsky,
Role of correlation and exchange for quasiparticle spectra of magnetic and diluted magnetic semiconductors.
\emph{Physica}   B \textbf{355}, 318-340  (2005);   [arXiv:cond-mat/0403266]. 
%
%
%
%
%
\bibitem{petr93}
J. Petru, Self-consistent treatment of the Anderson model.
\emph{Z. Physik} B \textbf{143}, 351 (1993).
%
%
%
\bibitem{wang99}
Hong-Gang Luo, Ju-Jian Ying and  Shun-Jin Wang, Equation of motion approach to the solution of the Anderson model.
\emph{Phys. Rev.} B \textbf{59}, 9710 (1999).  
%
% 
%
\bibitem{schaf99}
S. Schafer and P. Schuck, 
Dyson equation approach to many-body Green's functions and self-consistent RPA: Application to the Hubbard model. 
\emph{Phys.Rev.} B \textbf{59}, 1712 (1999). 
%
%
%
%
\bibitem{alas04} A. T. Alastalo, M. P. V. Stenberg and M. M. Salomaa, Response functions of an artificial atom in 
the atomic limit.
\emph{J. of Low Temp. Phys.}   \textbf{134}, 897 (2004).   
%  
%
%
%
%
\bibitem{kogo06}
I. Kogoutiouk  and H. Terletska,   
Investigation of the density of states in the non-half-filled two band periodic Anderson model. 
\emph{Int. J. Mod. Phys.} B \textbf{20}, 3101  (2006).
%
%
%
%
\bibitem{gors13}
G. Gorski and J. Mizia, 
Antiferromagnetic ordering of itinerant systems with correlation and binary disorder.
\emph{Physica} B \textbf{409}, 71 (2013). 
%
%
%
%
%
\bibitem{miz13}
G. Gorski and J. Mizia, 
Equation of motion solutions to Hubbard model retaining Kondo effect.
\emph{Physica} B \textbf{427}, 42-46 (2013).
%
%
%
%
%
%
\bibitem{miz14}
G. Gorski, J. Mizia and K. Kucab, 
Alternative equation of motion approach to the single-impurity Anderson model.
arXiv:1404.4439 [cond-mat.str-el] (2014). 
%
%
%
%
\bibitem{gors14}
G. Gorski, J. Mizia and K. Kucab, 
Alternative equation of motion approach to the single-impurity Anderson model.
\emph{Acta Phys. Pol.} A \textbf{126}, No.4A,  A97 (2014).
% 
%
%
\bibitem{ashok15}
Ch. Narasimha Raju and  A. Chatterjee. 
Ground state energy, binding energy and the impurity specific heat of Anderson-Holstein model.
\emph{Canadian Journal of Physics}, 10.1139/cjp-2014-0432 (2015).   
%
%
%
%
%
%
%
\end{thebibliography}
\end{document}